\documentclass[preprint]{aastex}

\newcommand{\tn}{T\,Tau\,N}
\newcommand{\ts}{T\,Tau\,S}
\newcommand{\tsa}{T\,Tau\,Sa}
\newcommand{\tsb}{T\,Tau\,Sb}
\newcommand{\brg}{Br$\gamma$}

\shorttitle{The circumstellar environment of T\,Tau\,S}
\shortauthors{Duch\^ene et al.}

\received{2005 January 31}
\begin{document}

\title{The circumstellar environment of T\,Tau\,S at high spatial and
spectral resolution}

\author{G. Duch\^ene\altaffilmark{1}}\affil{Laboratoire
  d'Astrophysique, Observatoire de Grenoble, BP 53, F-38041 Grenoble
  cedex 9, France} \email{gaspard.duchene@obs.ujf-grenoble.fr}
  \author{A. M. Ghez, C. McCabe} \affil{Department of Physics and
  Astronomy, UCLA, Los Angeles, CA 90095-1562, USA}
  \author{C. Ceccarelli} \affil{Laboratoire d'Astrophysique,
  Observatoire de Grenoble, BP 53, F-38041 Grenoble cedex 9, France}
  \altaffiltext{1}{Department of Physics and Astronomy, UCLA}

\begin{abstract}
  We have obtained the first high spatial (0\farcs05) and spectral
  ($R\sim35000$) resolution 2\,$\mu$m spectrum of the \ts\ tight
  binary system using adaptive optics on the Keck~II telescope. We
  have also obtained the first 3.8 and 4.7\,$\mu$m images that resolve
  the three components of the T\,Tau multiple system, as well as new
  1.6 and 2.2\,$\mu$m images. Together with its very red near-infrared
  colors, the spectrum of \tsb\ shows that this T\,Tauri star is
  extincted by a roughly constant extinction of $A_V\sim$15\,mag,
  which is probably the 0\farcs7$\times$0\farcs5 circumbinary
  structure recently observed in absorption in the ultraviolet. \tsa,
  which is also observed through this screen and is actively
  accreting, further possesses a small edge-on disk that is evidenced
  by warm (390\,K), narrow overtone CO rovibrational absorption
  features in our spectrum. We find that \tsa\ is most likely an
  intermediate-mass star surrounded by a semi-transparent
  2--3\,AU-radius disk whose asymmetries and short Keplerian rotation
  explain the large photometric variability of the source on
  relatively short timescales. We also show that molecular hydrogen
  emission exclusively arises from the gas that surrounds \ts\ and
  that its spatial and kinematic structure, while providing suggestive
  evidence for a jet-like structure, is highly complex.
\end{abstract}

\keywords{stars: individual (T\,Tau) --- binaries: close --- stars:
pre-main sequence --- circumstellar matter}


\section{Introduction}

Since its discovery by Dyck, Simon \& Zuckerman (1982), the companion
to the prototypical low-mass young stellar object T\,Tau has been the
subject of numerous studies and a source of many debates. In short,
while \ts\ has never been detected at optical wavelengths (Gorham et
al. 1992; Stapelfeldt et al. 1998a), it is much brighter in the
mid-infrared than the optically bright T\,Tau (Ghez et al. 1991),
which we will refer to as \tn\ in the following. This peculiar
characteristic is typical of a handful of other objects, collectively
known as infrared companions (IRC, Koresko et al. 1997). Noticeably,
the bolometric luminosity of \ts\ (12\,$L_\odot$, Koresko et al. 1997)
is larger than that of the 2\,$M_\odot$ \tn. Furthermore, while \tn\
has remained remarkably stable in recent years after significant
historical variations (Beck \& Simon 2001; Beck et al. 2004), large
photometric variations have been observed in \ts\ (Ghez et al. 1991;
Beck et al. 2004). \ts\ has also long been known as the source of
variable, polarized gyrosynchrotron radio emission (Skinner \& Brown
1994).

Studying the spectral energy distribution (SED) of \ts, Koresko et
al. (1997) found a bolometric temperature of $\sim500$\,K, much lower
than that of a normal T\,Tauri star (TTS) and typical of more deeply
embedded Class\,I protostars. However, the close vicinity and assumed
coevality with \tn\ appears to rule out the possibility of \ts\ being
in a much less evolved stage. Koresko et al. therefore proposed that
\ts\ is a normal TTS, possibly slightly more massive than \tn, that is
embedded within a compact opaque envelope responsible for a (variable)
$A_V\sim35$\,mag. If its envelope is not larger than a few 100\,AU,
then one can argue that this is a different, later, evolutionary stage
than Class\,I protostars, whose envelopes typically extend over
several thousand AU (Motte, Andr\'e \& Neri 1998; Motte \& Andr\'e
2001).

Ghez et al. (1991), van den Ancker (1999) and Beck et al. (2004) have
used the depth of the 10\,$\mu$m silicate and 3\,$\mu$m ice absorption
features in the spectrum of \ts\ to estimate an extinction ranging
from $A_V=5$ to 30\,mag with significant variations over time. For
comparison, the extinction to \tn\ is estimated to be
$A_V\sim1.4$--1.5\,mag (Kenyon \& Hartman 1995; Koresko et
al. 1997). Such large extinctions to \tsa\ explain its non-detection
in the optical. Beck et al. (2004) argued that the photometric
variability of \ts\ can be explained by large extinction changes in
our line of sight. Hogerheijde et al. (1997) have proposed that the
obscuring material in front of \ts\ is the almost pole-on
circumstellar disk of \tn. However, Akeson et al. (1998) concluded
that the outer radius of \tn's disk is only $\sim40$\,AU, about half
the \tn--\ts\ projected separation.

Another interpretation of the large variability of \ts\ is that it is
in fact a FU\,Ori-like object in which the luminosity is entirely
dominated by emission from the accretion disk (Ghez et
al. 1991). Variability then arises from changes in the accretion rate
on the star and the strong brightening observed in the early 1990s can
be interpreted as an accretion outburst. Ghez et al.'s claim was in
part based on the fact that the large flare they observed was
essentially grey over the 2--10\,$\mu$m range. More recently, Beck et
al. (2004) found that the flux of the Br$\gamma$ emission line, which
is thought to be powered by accretion, correlates with the continuum
flux in a way that qualitatively agrees with this scenario. However,
Beck et al. also found significant color variability for \ts,
seemingly contradicting the FU\,Ori scenario.

To further complicate the picture, Koresko (2000) used speckle
interferometry to identify a close ($\sim$0\farcs05 or 7\,AU at a
140\,pc distance, Bertout et al. 1999) companion to the IRC. Such a
tight separation immediately made \ts\ one of the most promising
system to determine a TTS dynamical mass. Its orbital motion has been
repeatedly monitored with near-infrared (K\"ohler et al. 2000;
Duch\^ene, Ghez \& McCabe 2002; Furlan et al. 2003; Beck et al. 2004)
and radio (Loinard et al. 2003; Johnston et al. 2003, 2004)
high-angular resolution techniques. Opposite claims have been made
regarding the possibility that the \tsa--\tsb\ system is physically
bound (Loinard et al. 2003; Furlan et al. 2003; Johnston et al. 2003,
2004; Beck et al. 2004). Although the centimeter radio observations
provide a $\sim$20~year time baseline, only \tsb\ is detected at these
wavelengths, requiring critical but uncertain assumptions to be
made. The near-infrared datasets, on the other hand, have spatially
resolved \ts\ for only the last few years. Yet, preliminary orbital
solutions based exclusively on the near-infrared images, which clearly
pinpoint all components of the system, suggests an orbital period on
order a few decades, a periastron separation of $\sim 5$--10\,AU at
most and a system mass of several solar masses (Beck et
al. 2004). Such a large mass suggests that \tsa\ is the most massive
component of the T\,Tau system although it remains unclear whether its
high dynamical mass can be readily reconciled with its moderate
luminosity (Johnston et al. 2003). Furthermore, the fact that the most
massive object has remained more deeply embedded than its lower mass
close companions is also against the natural expectation of a faster
clearing timescale for circumstellar material with increasing stellar
mass.

From a spatially resolved moderate resolution near-infrared spectrum,
\tsb\ has been found to be a normal low-mass TTS of spectral type
$\sim$M1 suffering from a significant extinction ($A_V\gtrsim8$\,mag,
Duch\^ene et al. 2002). This spectrum also shows \tsa\ to have a
featureless continuum with a strong Br$\gamma$ emission line
synonymous of accretion. The binarity of \ts, combined with its
strongly variable flux ratio (see Sect. 3.1; Beck et al. 2004) raises
serious concern about the conclusions that can be derived from the
unresolved near-infrared properties of the system. Clearly, one needs
to spatially resolve the tight binary system to study the variability
of both components in order to reach firm conclusions regarding their
nature, since the brighter component at $K$ has changed over time.

Some of the peculiar properties of \tsa, namely its featureless $K$
band spectrum and extremely red colors, do not apply to \tsb, which
appears to be a normal TTS (Duch\^ene et al. 2002). Therefore, the IRC
phenomenon in the T\,Tau system is limited to \tsa, and has to be
contained within a volume of a few AU, even though some extinction is
present in front of both components. A compact and opaque envelope
could lie around that source and reprocess stellar starlight at
mid-infrared wavelengths, and thereby erase any photospheric feature in
its spectrum, as suggested by Koresko et al. (1997). However, the
dynamical timescale for this compact envelope (free-fall time scale of
a few years) to be accreted and/or dispersed is several orders of
magnitude shorter than the estimated age of the system ($\sim1$\,Myr,
White \& Ghez 2001). An alternative scenario to explain the apparent
faintness of \tsa\ consists of invoking a small edge-on disk
circumstellar disk. This scenario, evoked by Hogerheijde et al.
(1997), Koresko (2000) and Beck et al.  (2004) appeared to gain ground
when Walter et al. (2003) detected an extended absorption feature that
is entirely opaque to ultraviolet photons at the location of
\ts. However, the size of this feature, $\sim$70\,AU, is much larger
than the few AU maximal size of a circum{\it stellar} disk around
\tsa. Until now the location of the extincting material in front of
\tsa\ remains subject to debate.

Besides large quantities of dusty obscuring material, the environment
of \ts\ also contains numerous evidence for mass-loss. A large and
complex CO outflow is known to arise from the whole T\,Tau system
(Edwards \& Snell 1982; Schuster, Harris \& Russell 1997). On smaller
scales, a multi-component dynamical structure has been discovered and
studied through optical emission lines by B\"ohm \& Solf (1994) and
Solf \& B\"ohm (1999). These authors have proposed that the T\,Tau
system is the source of two distinct stellar jets: \tn\ would drive an
East-West jet that almost points to the observer whereas \ts\ would be
the source for a North-South jet that extends in the plane of the
sky. It is not known whether this second jet comes from \tsa\ or \tsb,
though. Imaging of the near-infrared shock-excited H$_2$ emission line
has revealed extended emission over several arcseconds (Herbst et
al. 1996). On subarcsecond scales, Herbst et al. and Beck et
al. (2004) detected emission arising from \ts\, but higher spatial
resolution studies by Duch\^ene et al. (2002) and Kasper et al. (2002)
found no significant emission at the stars' location. This led Beck et
al. (2004) to suggest that the line emission arises only from shocked
gas in the immediate surroundings of the two stars, in area small
enough to be unresolved in direct imaging studies with resolution
$\sim$1\arcsec, but large enough to be resolved out in higher angular
studies. Using narrow-band high angular resolution images, Quirrenbach
\& Zinnecker (1997) argued that some of the complex spatially resolved
H$_2$ emission may arise from a different structure than a stellar
outflow. Stapelfeldt et al. (1998a) attempted to model the scattered
light images of the reflection nebula centered on T\,Tau, which likely
traces an outflow cavity, and found an ``intermediate'' inclination
angle, in disagreement with the inclination of either jet identified
by Solf \& B\"ohm (1999). Overall, while mass loss clearly occurs in
the environment of T\,Tau with at least two distinct outflows, our
understanding of the actual source and physical properties of both
jets is still poor.

In this paper, we present the first high spectral resolution
($R\sim35000$), high spatial resolution (0\farcs05) study of the \ts\
tight binary system. With this dataset, we have 1) detected
accretion-induced Br$\gamma$ emission from both \tsa\ and \tsb, 2)
shown that the H$_2$ emission is exclusively associated to the
surrounding gas and has a highly complex kinematic and spatial
structure, 3) detected $^{12}$CO absorption features from warm gas in
the line of sight of \tsa, and 4) measured radial velocities for both
components of the system. We also obtained new high angular resolution
1.6 and 2.2\,$\mu$m images of the system, as well as the first 3.8 and
4.7\,$\mu$m images that spatially resolve all three components of
T\,Tau. Combining the $^{12}$CO absorption lines with a new multicolor
photometric analysis of \tsb, we show that \tsa\ is an
intermediate-mass star surrounded by a small, semi transparent,
edge-on disk and that a wider, possibly circumbinary, structure is
located in front of both components of \ts.

The plan of this paper is as follows: in Sect.~2, we present the
imaging and spectroscopic datasets used in this study and their
reduction procedures. In Sect.~3 and 4, we present we present our
photometric and spectroscopic results, respectively. We discuss them
in Sect. 5 and summarize our findings in Sect.~6.


\section{Observations}

All observations used in this study have been obtained using the
adaptive optics (AO) system on the 10\,m Keck~II telescope (Wizinowich
et al. 2000). We have obtained direct images from 1.6\,$\mu$m to
4.7\,$\mu$m that clearly resolved the three components of the system
(Sect.~2.1), as well as spatially resolved high spectral resolution
spectra at 2\,$\mu$m of \tsa\ and \tsb\ (Sect.~2.2). Here we present
the instrumental setups used for these observations as well as the
data reduction processes. In all observations, the optically bright
\tn\ ($R=9.2$) was used as a natural guide star for the AO system.


\subsection{Near-Infrared Imaging}

On 2002 December 13, we used the facility AO-dedicated near-infrared
camera NIRC2 (Matthews et al. 2005, in prep.), a 1024$\times$1024
detector for which we used the 0\farcs00995$\pm$0\farcs00005 pixel
scale (Ghez et al. 2004) and whose absolute orientation on the sky is
0\fdg7$\pm$0\fdg2 (Beck et al. 2004). To avoid saturation of the
stars in the 2\,$\mu$m band, we used two narrow band ($\sim1.5\,\%$
bandpass) filters centered at 2.18\,$\mu$m (Br$\gamma$) and
2.28\,$\mu$m ($K_{cont}$), respectively. We further obtained for the
first time broadband AO images of the system at 3.8\,$\mu$m ($L'$) and
4.7\,$\mu$m ($M_s$). For all observations, we used the ``inscribed
circle'' pupil mask that decreases the amount of background emission
at the cost of a slight loss of resolution (the equivalent telescope
diameter is 9\,m). In each filter, we used a similar sequence in which
the system was placed in 2 or 4 different locations on the chip and
100 short integrations (0.02--0.05\,s) where coadded at each
location. Shorter integrations, obtained by reading only a fraction of
the array (256$\times$264 at $L'$ and 128$\times$152 at $M_s$), were
used at longer wavelengths to prevent saturation in the background.

On 2003 December 12, we obtained additional images of the system using
the near-infrared imager and spectrograph NIRSPEC (McLean et
al. 2000). Its 256$\times$256 imaging detector provides a pixel scale
of 0\farcs0168$\pm$0\farcs0001 and its absolute orientation on the sky
is 1\fdg1$\pm$0\fdg8, as estimated by us from images of several
well-studied calibration binaries. We used the $K$ band (2.2\,$\mu$m)
filter as well as an instrument-specific 1.6\,$\mu$m filter that is
very similar to the standard $H$ band filter. The same observing
technique was used as during the NIRC2 observations described above.

\begin{figure}
%
%
\epsscale{0.75}
\plotone{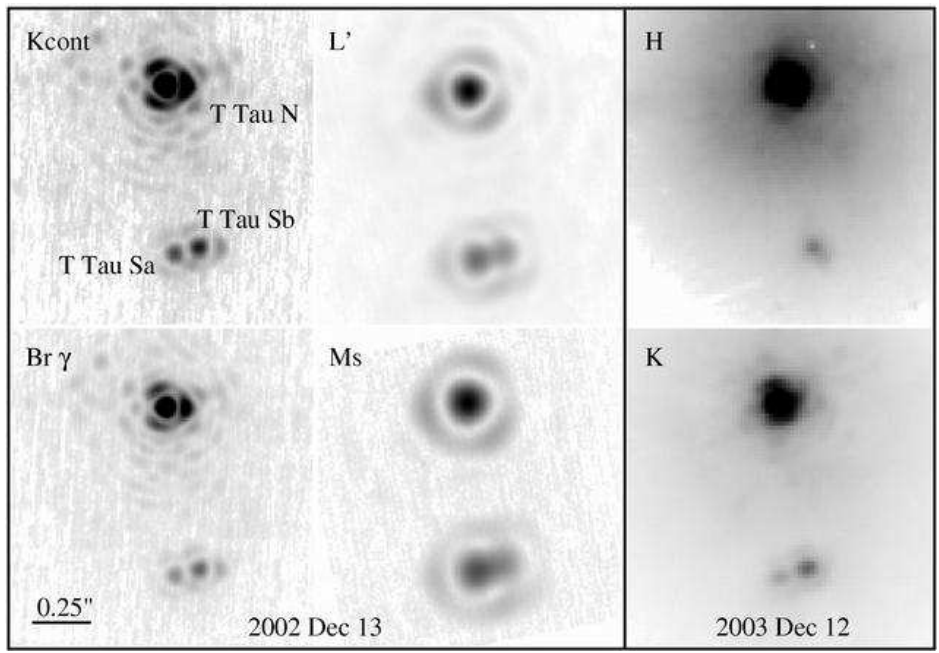}
\caption{Keck adaptive optics images of T\,Tau, presented on a square
  root stretch. All images are 1\farcs3 on the side, oriented with
  North up and East to the left. Note that \tsa\ was much redder than
  \tsb\ in December 2002 and was undetected in our December 2003
  $H$-band image.\label{fig:images}}
\end{figure}

All datasets were sky subtracted, flat-fielded, corrected for bad
pixel and shift-and-added using standard IRAF\footnote{IRAF is
distributed by the National Optical Astronomy Observatories, which is
operated by the Associations of Universities for Research in
Astronomy, Inc., under contract to the National Science Foundation.}
routines. The final images are presented in
Figure\,\ref{fig:images}. The spatial resolution, as measured from the
FWHM of \tn\ in all images, is 0\farcs055 , 0\farcs072, 0\farcs087 and
0\farcs107 at 2.2, 1.6, 3.8 and 4.7\,$\mu$m, respectively. All three
components are clearly detected in all images, except for the $H$ band
image in which \tsa\ is not detected. The two components of the
$\sim$0\farcs1 {\ts} binary are only barely resolved at
$M'$. Nonetheless, it is still possible to accurately extract the
astrometric and photometric properties of the binary.

For all images, we performed point spread function (PSF) fitting using
{\tn} as a perfect nearby and simultaneous PSF. Uncertainties on the
stars' locations and fluxes were determined from the standard
deviation of the quantities obtained in each individual image. All
simultaneous images yielded fully consistent astrometric measurements
for both the {\tn}--{\tsa} and {\tsa}--{\tsb} pairs. The astrometric
uncertainties for the former, wider, pair are dominated by the
detectors' calibration uncertainties whereas centroiding uncertainties
lead to the dominant source of error regarding the former, tighter,
binary. The astrometric results are summarized in
Table\,\ref{tab:astrom} and the photometric results are given in
Table\,\ref{tab:phot}. The average astrometric measurements from all
filters are indicated for each run.


\subsection{High-Spectral Resolution Spectroscopy}


\subsubsection{Instrumental setups and datasets}

\begin{figure}
\epsscale{1.0}
\plotone{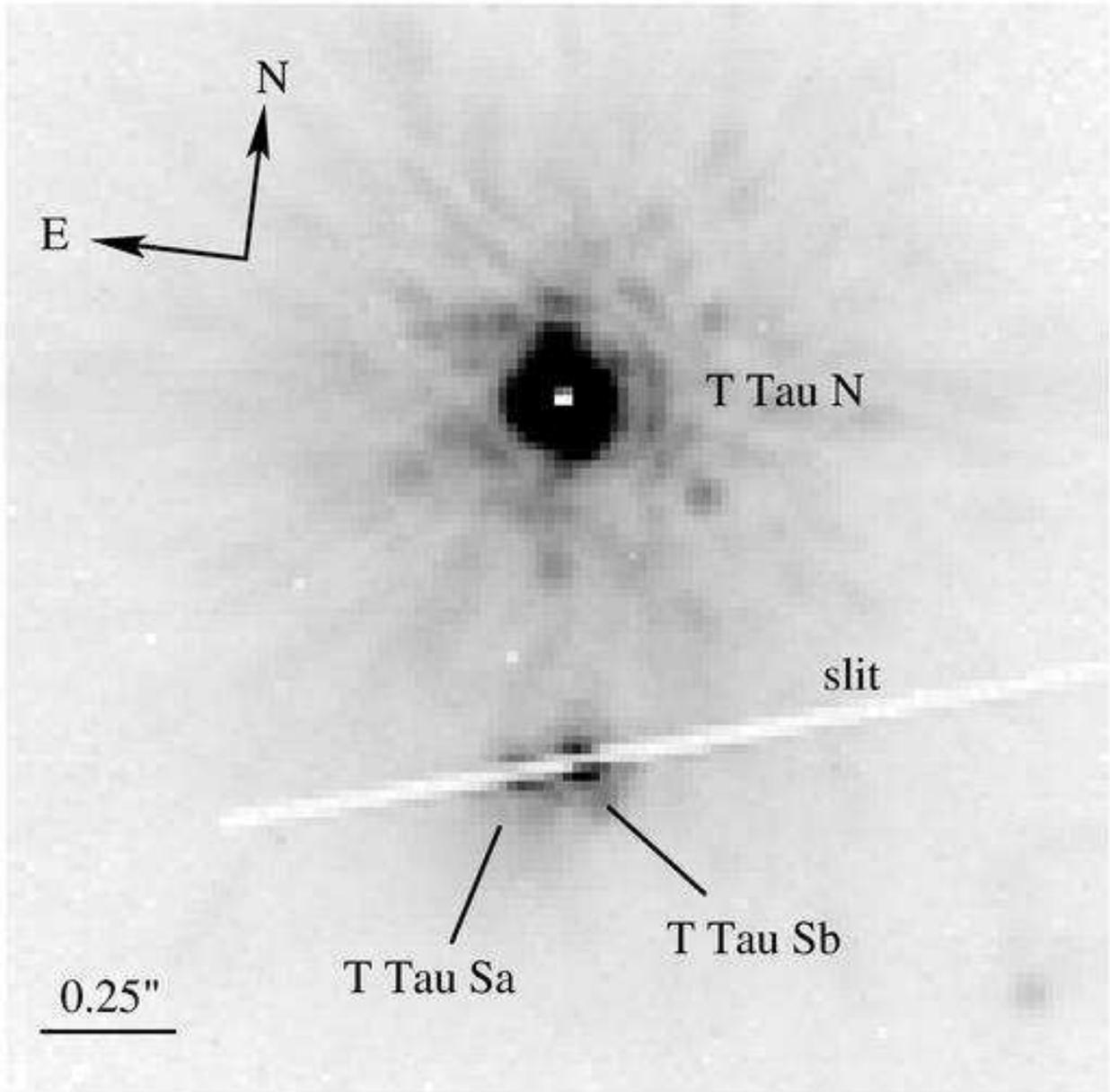}
\caption{$K$-band image of the T\,Tau triple system obtained with the
  slit-viewing camera of NIRSPEC during the acquisition of the
  high-resolution spectra presented in this study. \tn\ is heavily
  saturated in this image. The image is 2\arcsec\ on the
  side.\label{fig:img_slit}}
\end{figure}

On 2003 December 12, we used NIRSPEC to obtain 2\,$\mu$m
cross-dispersed high resolution spectra of the \ts\ binary system. As
illustrated in Figure\,\ref{fig:img_slit}, we aligned the
0\farcs027$\times$2\farcs26 slit along the binary system so as to
spatially disentangle the spectra of the two components. In order to
obtain an absolute calibration of the radial velocities, we also
observed HD\,35410, an IAU radial velocity standard (spectral type
G9III, $v_{rad}=20.3$\,km~s$^{-1}$, Fehrenbach \& Duflot
1980). Observations of the early A-type stars HR\,1389 and HD\,36150
were made with the same set-up immediately before or after \ts\
and radial velocity standard in order to measure the telluric
transmission function. Each star was located at 2 to 6 different
locations behind the slit to correct for bad pixels and increase the
signal-to-noise ratio without saturating the stars. We obtained total
integration times of 30\,min for \ts\ and ranging from 40 to 120\,s
for HR\,1389, HD\,35410 and HD\,36150.

We selected a single cross-disperser set-up that allowed us to place
simultaneously 7 consecutive, though non-adjacent, $K$ band orders on
the 1024$\times$1024 spectroscopic detector for all objects. Of these,
the first two correspond to the short-wavelength end of the $K$ band,
where strong telluric absorption features are located. The
non-simultaneousness of the observations of the targets and their
calibration A-type star resulted in poor telluric corrections in these
two orders and they are not discussed further. The wavelength range of
the remaining five orders are indicated in Table\,\ref{tab:setup}. Our
spectral coverage includes molecular (2.1218 and 2.2477\,$\mu$m) and
atomic (2.1661\,$\mu$m) hydrogen lines, as well as three of the four
$\Delta v=2$ rovibrational $^{12}$CO bandheads (2.293, 2.322 and
2.383\,$\mu$m), as well as a slew of associated individual $^{12}$CO
transitions.

In addition to obtaining standard calibration datasets (halogen flat
field lamp and adjacent sky spectrum), we obtained spectra of arc
lamps that were used to determine the wavelength solution for our
dataset (see Sect~2.2.3). From the measured width of unresolved arc
lines, we estimate that the spectral resolution is on order
$R\sim35000$ (8.5 km~s$^{-1}$), nominal for the 2-pixel wide slit
used in our set-up. We also obtained spectra of the Fabry-Perrot
``etalon'' lamp which produces a spectrum with $\sim15$ emission lines
equally spaced in frequency in each order.


\subsubsection{Data Reduction and Spectra Extraction}

All steps in the data reduction were performed with IRAF tasks. The
raw spectra were first sky-subtracted using a sky spectrum or, if it
yielded too large or too small a residual background level, another
spectrum of the same object with the star(s) at a different
location. A constant value per detector quadrant was then removed to
correct for small bias differences in the four amplifiers. The spectra
were flat-fielded using the normalized difference between the ``on''
and ``off'' halogen lamp spectra and finally corrected for bad pixels
and cosmic rays. Once these cosmetic corrections were performed, we
separated each order in the spectrum using windows that encompass each
of them with all of their corresponding sky area. We then corrected
each order for distortion using the spectra of bright stars and of
the etalon lamp lines to track the curvature of the spatial and
spectral axes, respectively. After the distortion correction, the
2-dimension spectra have the spatial axis vertical and the spectral
axis horizontal.

The spectra of single stars, such as the radial velocity standard or
the A-type stars used for atmospheric corrections, were extracted
using a $\sim$0\farcs4 aperture centered on the star. For \ts, this
was not practical as the spectra of the primary overlaps significantly
on the position of the secondary and vice-versa. Both stars must be
carefully deblended to obtain uncontaminated spectra. We used a
custom-made routine that, for each pixel along the spectral axis of
each order, fits two Gaussian profiles of equal width along the
spatial axis. Because the core of the AO-produced PSF is not a perfect
Gaussian, this fit is not perfect and some contamination of one
component by the other is unavoidable. We find that, depending on
wavelength and AO correction, the spectrum of the fainter component
(\tsa) can be under- or over-estimated by up to 10\,\% although in
most cases this effect does not exceed a few percent. Lorentzian
profiles were also tested but they result in noisier and more
contaminated spectra.

To correct the extracted 1-dimensional spectra for telluric
absorption, we used the spectra of the A-type stars, which
intrinsically contain essentially no photospheric line, except for the
HI Br$\gamma$ line (in order \#2) which is spectrally resolved over
$\gtrsim$100 pixels given the typical rotational velocities of such
objects ($\sim200$\,km~s$^{-1}$). The many narrow features observed in
the spectra of the A-type stars are therefore unresolved telluric
absorption features. We fitted a 20--30$^{th}$ order polynomial
function to the continuum (and hydrogen line) across each order and
divided the observed spectrum by this ``continuum''. This yielded the
telluric transmission function, which we normalized at unity between
absorption lines. The spectra of \ts\ and HD\,35410 were then divided,
order by order, by their corresponding telluric transmission function
to yield the final spectra used in our analysis.

The final spectra have signal-to-noise ratios on order 60--90 for
\tsa\ and 90--120 for both \tsb\ and HD\,35410, as estimated from the
standard deviation of the counts within several featureless continuum
windows.


\subsubsection{Wavelength Calibration}

Too few afterglow OH lines were present even in our 300\,s individual
exposures to derive an accurate wavelength solution. We therefore
relied on the identification of a few (between 2 and 7) known arc
lines in each order. We used each independent pair of lines to
estimate the pixel size (in km~s$^{-1}$) in separate sections of all
orders. We found no significant systematic variations at the
$\sim3$\,\% level within or between the orders. The average spectral
pixel size for each order are listed in Table\,\ref{tab:setup}; the
average and standard deviation over all orders is $<\delta
v>=4$.31$\pm$0.11\,km~s$^{-1}$\,/pixel.

For each order, the wavelength solution was calculated as $\lambda_n =
\lambda_i \times (1 + n\ \delta v/c)$ where $n$ is the running pixel
number starting at 0. The values of $\lambda_i$ listed in
Table\,\ref{tab:setup} are the average values over all identified
lines in each order. Since orders \#3 and \#5 contains the largest
number of identified arc lines (4 and 7, respectively), we used the
dispersion of $\lambda_i$ to estimate that the accuracy of our
absolute wavelength calibration is on order 15\,km~s$^{-1}$; this is
dominated by uncertainties on the exact rest wavelength of the arc
lamp lines and on the approximative nature of the adopted wavelength
solution. For all other orders, we assume that the uncertainty is the
same. The frequency difference between any two consecutive etalon lamp
lines was found to be constant throughout the orders at the 2.5\,\%
level, confirming the validity of our wavelength solution.

Given the uncertainty in our absolute wavelength calibration, it is
not possible to determine radial velocities directly from the measured
wavelength of spectral features. Rather, we used cross-correlation
with the spectrum of HD\,35410, which takes full advantage of the many
photospheric features located in the $K$ band. Using entire orders in
the cross-correlations yields relative pixel shifts that are accurate
to within 0.2 pixel ($\lesssim1.0$\,km~s$^{-1}$, see Sect.~4.5). We
then transformed the pixel shifts into velocity shifts using the
spectral pixel size defined above and, correcting for the Earth motion
using the IRAF task {\it rvcorr} (the correction is on order 6.5
km~s$^{-1}$), we derived the heliocentric radial velocity of each
source. Since the radial velocity of HD\,35410 is known to within
0.7\,km~s$^{-1}$ (Fehrenbach \& Duflot 1980), the cross-correlation
technique therefore provides an accurate estimate of the radial
velocity of the components of \ts.


\section{Imaging Results}


\subsection{Photometric Variability}

{\tn} has not shown signs of near-infrared variability over the last
decade (Beck et al. 2004), and we assumed that it was at its stable
level in all bands at the time of our observations\footnote{The same
magnitude was assumed for both the $K_{cont}$ and \brg\ narrow band
filters as for the broad band $K$ filter.}. We therefore adopt the
following photometry for {\tn}: $M_s=2.95$, $L'_{TTN}=4.32$,
$K_{TTN}=5.52$ and $H_{TTN}=6.32$ (Ghez et al. 1991; Beck et
al. 2004). With this simultaneous photometric reference, we can study
the variability of both \tsa\ and \tsb.

\begin{figure}
\epsscale{1.0}
\plotone{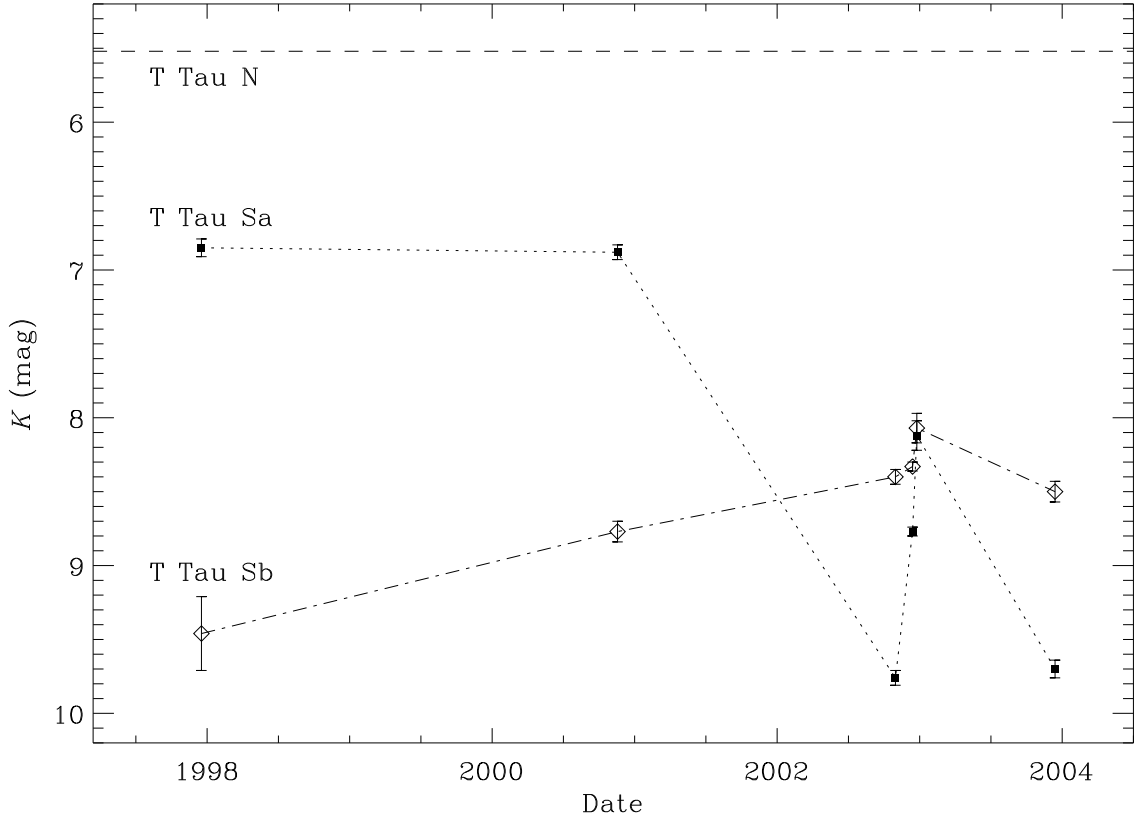}
\caption{The spatially-resolved time variability of both components of
  \ts\ in the $K$ band from Koresko (2000), Duch\^ene et al. (2002),
  Furlan et al. (2003), Beck et al. (2004) and this study. Filled
  squares and open diamonds represent \tsa\ and \tsb\ respectively;
  the dotted and dash-dotted lines should only be considered as
  guidelines for the eye. The dashed line represent the constant
  brightness of \tn\ (Beck et al. 2004). Uncertainties of 0.05\,mag,
  0.05\,mag and 0.10\,mag for the \tn--\ts\ flux ratios have been
  assumed for the Koresko (2000), Beck et al. (2004) and Furlan et
  al. (2003) measurements, respectively. The dramatic flux decrease of
  \tsa\ between late 2000 and now has been accompanied by only minor
  changes in the flux of \tsb.\label{fig:var}}
\end{figure}

The images presented in Figure\,\ref{fig:images} show that, at the
time of both our observations, \tsa\ was the faintest component of the
T\,Tau triple system at 1.6 and 2.2\,$\mu$m, in contrast to all
spatially resolved measurements of the system prior to 2002 (Koresko
2000; Duch\^ene et al. 2002). As shown in Figure\,\ref{fig:var}, \tsa\
is now $\sim$3\,mag fainter ($K\sim9.8$) and much redder ($H-K>4.9$)
than it was at the time of the first images that spatially resolved
the triple system ($K\sim6.9$ and $H-K\sim2.6$ in November 2000,
Duch\^ene et al. 2002). In late 2002, \tsa\ brightened by
1.7\,magnitude at $K$ band over only 2 months, indicating large
variability on short timescale. At longer wavelengths, we also find
\tsa\ to be fainter than \tn, and the entire \ts\ binary system has
reached its lowest $L'$ flux since Dyck et al. (1982)'s discovery
observations ($L'_{TTS}=5.15$; see Beck et al. 2004). At $M_s$, the
combined \ts\ system was 0.4\,mag fainter than \tn, while all
historical measurements found \ts\ to clearly dominate the flux from
the system at 4.7\,$\mu$m (Ghez et al. 1991; Herbst et al. 1997). The
possible physical cause(s) of this variability will be discussed in
Sect.~5.

While \tsa\ is undergoing a rapid dimming trend since late 2000 (Beck
et al. 2004), \tsb\ has varied in a much more moderate way (see
Figure\,\ref{fig:var}). Its $K$ band brightness has increased from
$K=9.4$ (Koresko 2000) to $K=8.8$ (Duch\^ene et al. 2002) between 1997
and 2000 and, since then, it has stabilized around
$K=8.4\pm0.1$. Between November 2000 and December 2003, it has also
somewhat brightened in $H$ band, from $H=10.7$ to $H=10.3$, so that
its $H-K$ color index has only changed by 0.2\,mag.

From this comparison, it is clear that the photometric behavior of
both components of \ts\ is dramatically different and cannot be
explained with the same model. Furthermore, since most historic
photometric measurements of the unresolved \ts\ system have yielded
$K<8$, it is likely that \tsa\ dominated most of these
measurements and that the strong historic variability of the system is
primarily related to \tsa, a conclusion also supported by the
astrometric motion of \ts\ (Beck et al. 2004). 


\subsection{Extinction Estimates For \tsb}

From a lower spectral resolution 2\,$\mu$m spectrum taken in November
2000, we had placed a lower limit of $A_V=8$\,mag to the extinction
toward our line of sight to \tsb, conservatively assuming that the
excess was negligible in the $H$ band (Duch\^ene et al. 2002). Since
Muzerolle et al. (2003) have shown that the spectrum of the
1.5--4\,$\mu$m excess in normal TTS is well approximated by that of a
$\sim1400$\,K blackbody, we can now obtain an accurate estimate of the
actual extinction to this source for this former epoch as well as for
the more recent observations presented here.

The method consists in using simultaneous photometry of \tsb\ in two
near-infrared bands ($H$ and $K$ or $K$ and $L'$ here), estimating the
continuum excess contribution (``veiling'') at one of these
wavelengths from a spectrum, and separating the stellar and excess
brightnesses using their known spectral dependence. Given its spectral
type, \tsb's photosphere is characterized by $T_{eff}=3800$\,K, and so
the relative contribution of the veiling over the photosphere
increases with wavelength, according to their respective Planck
functions. To illustrate the method, we first revisit our extinction
estimate for the Nov 2000 dataset. The observed $K$ band veiling at
that time, $r_K\sim2$, implied an $H$ band excess on order
$r_H\sim0.7$. In turn, this means that, while the observed color index
was $H-K=1.93$, the excess-free color index was $H-K=1.13$. With an
expected photospheric color of $(H-K)_0=0.18$ (e.g., Kenyon \&
Hartmann 1995), this implies that $A_V = 15\pm3$\,mag assuming the
Rieke \& Lebofsky (1985) extinction law. The uncertainty was estimated
through the quadratic sum of typical uncertainties for each parameters
used in this calculation: $\sigma_{r_K} \sim 0.5$,
$\sigma_{T_{excess}} \sim 100$\,K, $\sigma_{T_{star}} \sim 100$\,K,
$\sigma(H-K)_{obs} \sim 0.1$, $\sigma(H-K)_0 \sim 0.02$ and
$\sigma((A_H-A_K)/A_V) \sim 0.005$.

Our Dec 2003 spectrum of \tsb\ shows that the 2\,$\mu$m excess was
also about $r_K\sim2$ (see Sect.~4.3). Using the same method as above,
we derive $A_V = 12\pm3$\,mag for this newer epoch at which we
obtained both $H$ and $K$ band fluxes for \tsb. In our Dec 2002
dataset, we can use the simultaneous $K$ and $L'$ band photometry of
\tsb, although the $L'$ band might be contaminated by the ice
absorption feature observed in the line of sight of the unresolved
\ts\ system (Beck et al. 2004). Keeping this caveat in mind and
assuming that the $K$ band veiling for \tsb\ was equal to those
measured in 2000 and 2003, we derive $A_V=17\pm4$\,mag.

The variations in extinction during these three epochs are not
significant, which is consistent with the moderate ($\lesssim
0.5$\,mag) variations of \tsb\ in both $H$ and $K$ bands. While this
method is not accurate enough to monitor the extinction toward \tsb,
it confirms that this component suffered only moderate changes in
extinction around an average value of $A_V\sim15$\,mag
while the brightness of \tsa\ in the $H$ and $K$ band dropped by
$>5$\,mag and about 3\,mag, respectively, over the 2000--2003 period.


\subsection{Orbital Motion}

Our December 2002 astrometric measurements presented here are fully
consistent with the October and December 2002 observations of Beck et
al. (2004). We find significant orbital motion within the \ts\ binary
system over the year separating our two datasets. Our most recent,
December 2003, astrometric measurement compares reasonably well with
the preliminary orbital solutions presented by Beck et al. (their
Figure~13), although it seems that \tsb\ is curving slightly more
outwards than their ``possible orbital models''. This is still within
the uncertainties of the orbital solution and does not prompt us to
revisit their results. Overall, as Beck et al. (2004) pointed out, it
is not possible yet to derive a full orbital solution only on the
basis of the handful of spatially resolved near-infrared datasets,
essentially because of the limited time coverage they offer. These
authors showed that only a strict lower limit of 1.6\,$M_\odot$ lower
limit to the total mass of the \ts\ system can be estimated for
now. 

The new astrometric measurements presented here have been taken about
six months before and after the latest published radio astrometric
measurement (Johnston et al. 2004). Our dataset can therefore help
better understanding the relationship between the radio source \ts\
and both infrared components of the tight binary system. In
particular, comparing the two approaches is critical as only one radio
source appears to be associated to \ts. It has been suggested by
Johnston et al. (2004) that this source, although physically
associated to \tsb\, may be displaced from it by 10--30\,mas. At the
time of the most recent radio observations (2003 June 23), the
separation between the \tn\ and \ts\ radio sources was
0\farcs684$\pm0$0\farcs006 at position angle 193\fdg2$\pm$0\fdg6
(Johnston et al. 2004). The average of the separations we measured in
December 2002 and December 2003 between \tn\ and \tsb\ is
0\farcs678$\pm$0\farcs005 at position angle 191\fdg8$\pm$0\fdg7. These
two separations are only different by about 1.5$\sigma$ and it is
therefore not possible to conclude about a possible systematic
difference between the radio and infrared sources.


\section{Spectroscopic Results}

In this section, we discuss the high resolution spectra obtained for
the \ts\ binary system. In studying the spatially resolved spectra of
\tsa\ and \tsb, we focus on the molecular (Sect.~4.1) and atomic
hydrogen (Sect.~4.2) emission lines, the photospheric features
(Sect.~4.3) and the gas-phase CO absorption line series
(Sect.~4.4). We also determine radial velocities for both components
of the system (Sect.~4.5).


\subsection{Molecular Hydrogen Emission}

\begin{figure}
\epsscale{1.0}
\plotone{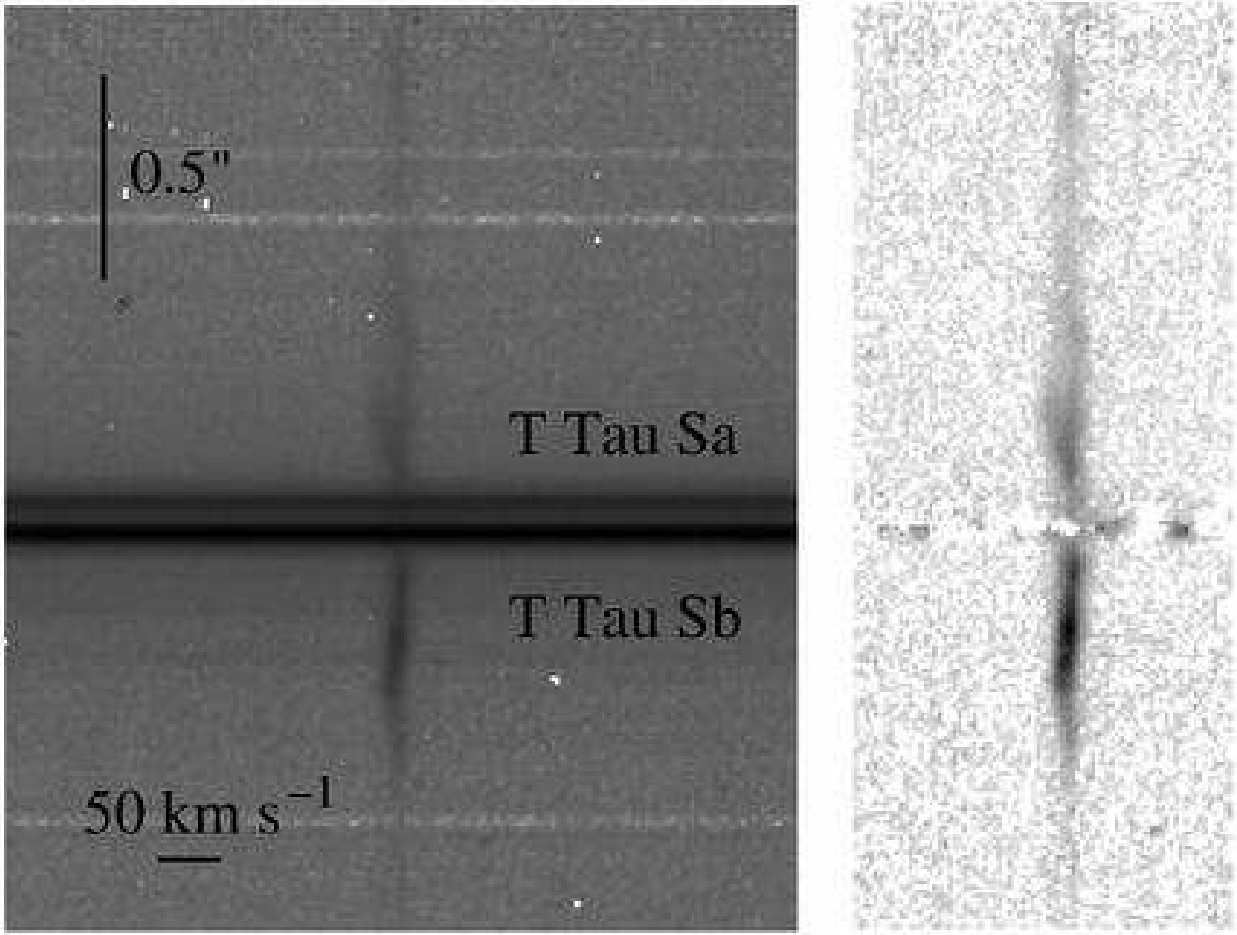}
\caption{\label{fig:2d_h2}{\it Left:} The 2D spectrum (3\farcs2 by
  700\,km~s$^{-1}$) of \ts\ around the H$_2$ emission line on a
  logarithmic stretch. The three negative stripes about 1\arcsec\
  from the stars are due to faint residuals in the sky image used in
  the data reduction process. Note that the spectrum extends over more
  than one slit length because we combined images from several slit
  positions. {\it Right:} The continuum subtracted image in a
  300\,km~s$^{-1}$-wide window around the H$_2$ emission line on a
  square root stretch. Note the complex velocity patterns on either
  side of the stars. Small residuals are left at the location of \tsb\
  because the spectrum of this object has higher frequency structure
  than the second-order polynomial continuum that we subtracted. No
  significant emission is detected at the location of either
  component.}
\end{figure}

Besides the spectra of both components, the 2-dimensional spectrum
corresponding to order \#1 reveals spatially extended emission in the
2.1218\,$\mu$m $v=1$--0\,$S$(1) H$_2$ line, as illustrated in
Figure\,\ref{fig:2d_h2}. This emission line, resolved on larger
spatial scales for this system by Herbst et al. 1996), is usually
detected in jets emanating from TTS and more embedded young stellar
objects and studying its spatial and velocity structure can provide
high spatial resolution information on the outflows in the vicinity of
\ts.

The spectra of both components show weak H$_2$ emission lines, with
equivalent widths (EWs) on order 0.4 and 0.15\,\AA\ for \tsa\ and
\tsb, respectively. The intensity of these lines, however, is not
larger than the emission from the gas in the vicinity of the stars and
it is possible that no significant emission is associated to the point
sources. To better isolate the line emission, we fitted and subtracted
a 2$^{\rm nd}$-order polynomial function representing the spectral
continuum at each pixel along the spatial direction. The result of
this subtraction is presented in the right panel of
Figure\,\ref{fig:2d_h2}, where only the line emission and small
residuals at the position of \tsb\ (which has a spectrum that contains
photospheric features which are not fitted here) are detected. There
is no detectable emission left at the location of either component,
providing a clear confirmation of the idea put forth by Beck et
al. (2004) that the line emission seen in lower spatial resolution
spectra was mistakenly attributed to the stars. Rather, it is clear
from our data that all the emission can be attributed to excited gas
in the vicinity of the binary system: this emission can be traced at
least over 1\arcsec\ on each side of the binary and the brightest
emission peak is offset by $\sim$0\farcs4 from \tsb.

\begin{figure}
%
%
\epsscale{1.0}
\plotone{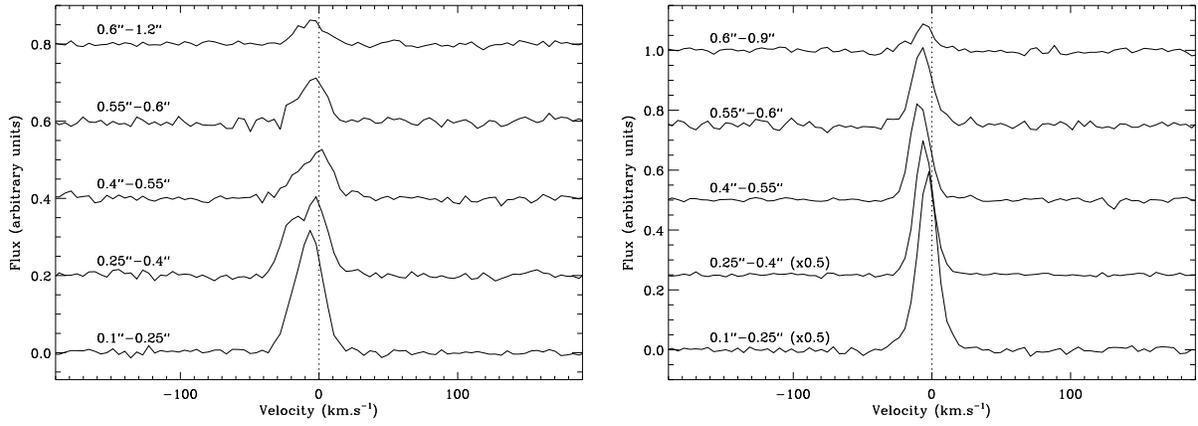}
\caption{Spectra of the H$_2$ emission line in successive windows
  located on the East ({\it left}) and West ({\it right}) side of the
  binary system. The rest velocity of \tsb\ is indicated by the dotted
  line.\label{fig:cuts_h2}}
\end{figure}

The kinematics of the gas surrounding the two stars can be analyzed
from the 2-dimensional spectrum (position-velocity diagram) shown in
Figure\,\ref{fig:2d_h2}. We have extracted the emission spectrum in
contiguous windows on both sides of the binary system; these spectra
are presented in Figure\,\ref{fig:cuts_h2}. In these spectra, we have
used the heliocentric radial velocity of \tsb\ (+21.1\,km~s$^{-1}$,
see Sect.~4.5) as zero velocity reference. The velocity structure of
the H$_2$ emission around \ts\ proves remarkably complex. On the
western side of the system, the line emission is seen to gradually
blueshift at increasing distance from the stars, by about
10\,km~s$^{-1}$ over $\sim$0\farcs5, before returning to the velocity
observed close to the stars. Several separate peaks of emission can be
identified along this direction. On the eastern side of the binary,
the line emission appears to fork into two components, one of which
becoming increasing bluer by about 10\,km~s$^{-1}$ over the first
0\farcs4, i.e. slightly faster than on the western side, before
gradually disappearing. The other component remains at a more or less
constants velocity over $\gtrsim$1\arcsec.

The 2.2477\,$\mu$m $v=2$--1\,$S$(1) H$_2$ line is included in order
\#3 in our dataset. However, we fail to detect it, indicating that it
is at least 17 times (3$\sigma$) fainter than the $v=1$--0\,$S$(1)
line on either side of \ts. On the other hand, the 2.2233\,$\mu$m
$v=1$--0\,$S$(0) line is located between orders \#1 and 2 and cannot
be studied here; we note, however, that this line was never detected
in this system however (e.g., Beck et al. 2004).

The spatial extension of the $v=1$--0\,$S$(1) emission line is fully
consistent with the findings of Herbst et al. (1996), but we now also
have access to kinematic information to study the structure of the gas
around \ts; we discuss this in Sect.~5.2. It is not immediately clear
how the H$_2$ emission we detect here relates to spatially resolved
emission from past studies. For instance, the northern component (N1)
of the fluorescent ultraviolet emission found by Saucedo et
al. (2003), which they interpret as arising from shock-excited regions
in the envelope of a wide-angle outflow, appears spatially coincident
with the infrared emission we detect to the west of the \ts\
system. On the other hand, the detailed studies of optical forbidden
line emission by Solf \& B\"ohm (1994) and B\"ohm \& Solf (1999)
revealed that two kinematic components (B and D, respectively
associated to the collimated E-W and N-S jet) could be spatially
connected with our observations at the West and East end of the slit,
respectively. The velocities they measure are much larger than those
derived here, but this would be a natural behavior due to the lower
excitation (e.g., lower velocity shocks) conditions required for the
H$_2$ emission line (e.g., Davis et al. 2001).


\subsection{Atomic Hydrogen Emission}

\begin{figure}
\epsscale{1.0}
\plotone{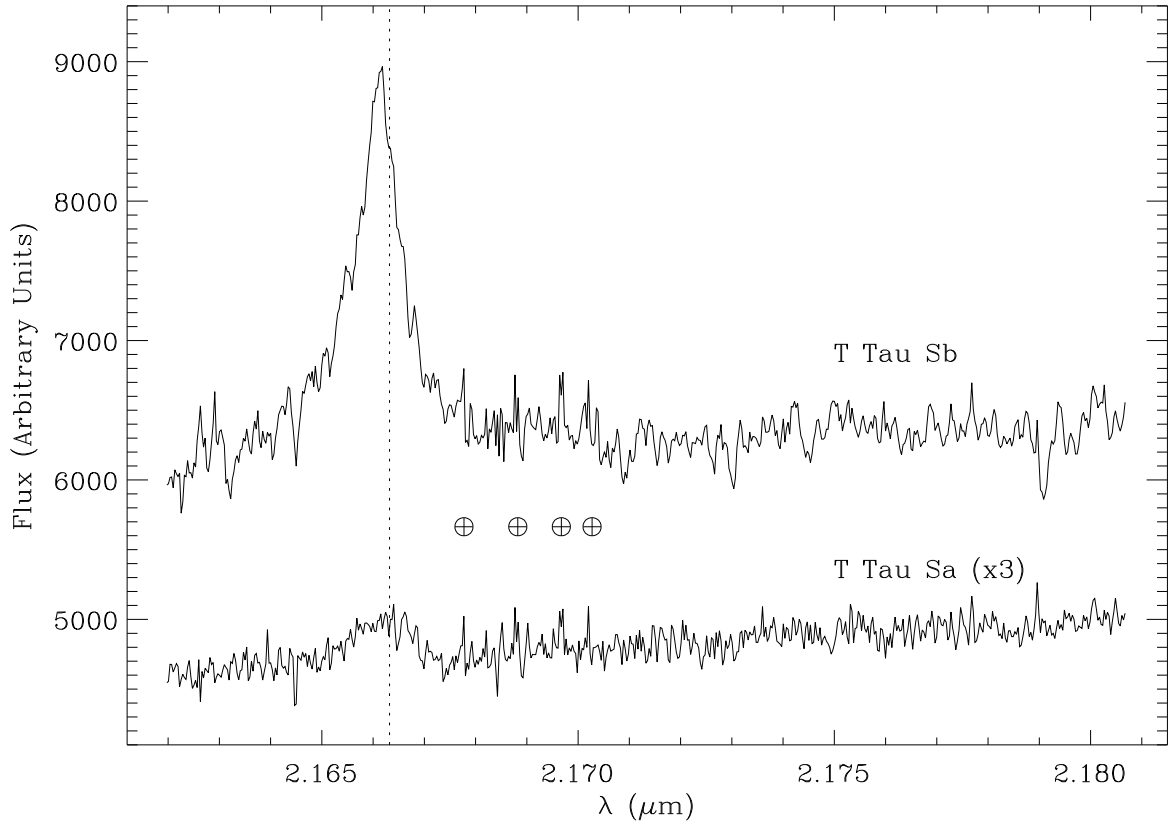}
\caption{The spectra of \tsa\ and \tsb\ around the Br$\gamma$ emission
  line. The vertical dotted line indicates the rest velocity of
  \tsb. A few spikes at wavelengths between 2.167 and 2.170$\,\mu$m
  (labelled with a $\oplus$ symbol) are due to an imperfect telluric
  absorption correction and are not intrinsic to the
  stars.\label{fig:spec_brg}}
\end{figure}

The spectra from order \#2 contains the atomic hydrogen Br$\gamma$
emission line, a tracer frequently used to study accretion in young
stellar objects based on its strength (Muzerolle et al. 1998) and
spectral profile (Folha \& Emerson 2001). As shown in
Figure\,\ref{fig:spec_brg}, both components show clear Br$\gamma$
emission, with \tsb\ displaying the strongest line. Despite its
weakness, the emission line in the spectrum of \tsa\ can be identified
even in single cuts along the spectra at the location of that
component and it cannot result from contamination by the strong
emission line of \tsb. As opposed to the molecular hydrogen emission
line, the Br$\gamma$ line is not spatially extended with respect to
the nearby continuum, showing that the emitting region is much smaller
than our spatial resolution of 0\farcs05.

The emission line properties are summarized in
Table\,\ref{tab:h_lines}. The Br$\gamma$ EW is 6.2 and 0.8\AA\ for
\tsb\ and \tsa, respectively. These values are typical of accreting
TTS (Muzerolle et al. 1998). While the EW of \tsb\ has only marginally
decreased with respect to our November 2000 observations (Duch\^ene et
al. 2002), the line strength has strongly decreased for \tsa. We are
not aware of other spatially resolved line strength measurements to
further analyze the amplitude of its variation on either
component. However, it must be emphasized that EWs is insensitive to
the line-of-sight extinction and therefore the observed variations for
\tsa\ can only be explained by an intrinsic change in line strength,
and therefore in accretion properties.

The line profiles are wide: their FWHM are $\sim$150\,km~s$^{-1}$ and
their 10\,\% (``zero'') full width at zero intensity (FWZI) range from
275 to 500\,km~s$^{-1}$. The profiles are roughly symmetrical, although
the blue side of the line has an EW 2.5--3 stronger than the red side
for both components. Furthermore, both emission lines extend to
significantly larger velocities in their red wing than in their blue
wing. Finally, the line peak for \tsb\ is significantly blueshifted
with respect to its rest velocity. From these properties, both objects
can therefore be classified as ``type I'' in the denomination of Folha
\& Emerson (2001). These authors have shown that this group is the
most frequent among TTS and the observed line profiles of \tsa\ and
\tsb\ are representative of normal TTS. There is little doubt that,
despite its unusual IRC classification, \tsa\ is accreting in a way
that is normal for a TTS. We note however that Davis et al. (2001)
found similar Br$\gamma$ emission line profiles for Class\,I
protostars so that the nature of \tsa\ cannot be firmly ascertained
from its Br$\gamma$ line profile.

Since the Br$\gamma$ line is assumed to be produced by gaseous
material accreting onto the star, it can be used to estimate its
accretion. Indeed, while the EW of the Br$\gamma$ line does not
correlate with accretion rate, its total flux does (Muzerolle et
al. 1998; Folha \& Emerson 2001). This is not practical for \tsa,
however, as the extinction towards this component is highly uncertain
(see Sect.~5.1), and we limit our analysis to \tsb. Combining the $K$
band flux with the line EW for this component, we derive an observed
(i.e., not extinction-corrected) line fluxes of
9.4$\times10^{-17}$\,W~m$^{-2}$. After correction for an extinction of
$A_V=15$\,mag (see Sect.~3.1), we estimate the accretion rate on \tsb\
to be on order 9.2$\times10^{-8}\,M_\odot$~yr$^{-1}$ (assuming a
2\,$R_\odot$ radius and 0.5\,$M_\odot$ mass). Using the Br$\gamma$ EW
and $K$ band fluxes measured in our November 2000 spectrum (Duch\^ene
et al. 2002) and the corresponding extinction derived above, we find
that the accretion rate for \tsb\ was essentially the same at that
time.

In summary, \tsb\ definitely appears as a normal though quite
extincted TTS with a roughly constant, relatively high
($\sim10^{-7}\,M_\odot$~yr$^{-1}$), accretion rate. On the other hand,
\tsa\ appears to be accreting at a probably variable, yet
undetermined, rate.


\subsection{Photospheric Features}

The 2\,$\mu$m spectral region contains several strong features that
are useful for determining the spectral type of late-type stars. Of
these, the 2.20\,$\mu$m Na doublet and 2.26\,$\mu$m Ca triplet are
unfortunately located in the gaps between the orders of our
cross-dispersed spectra and we cannot attempt to detect them. However,
orders \#4 and \#5 contains three overtone ($\Delta v=2$)
rovibrational $^{12}$CO bandheads and numerous associated individual
transitions that are prominent in late-type stars.

\begin{figure}
%
%
\epsscale{1.0}
\plotone{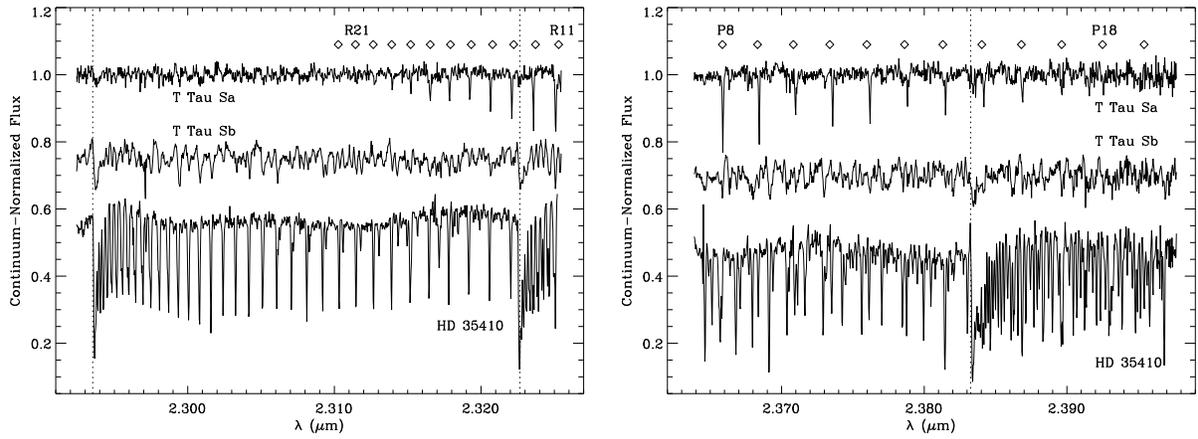}
\caption{The spectra of \tsa, \tsb, and HD\,35410, from top to
  bottom,in orders \#4 ({\it left}) and \#5 ({\it right}). They all
  have been continuum-normalized and shifted vertically by different
  amounts in the two orders for clarity. Vertical dotted lines
  indicate the location of the $^{12}$CO bandheads while diamonds
  indicate $v=2$--0 $^{12}$CO individual transitions; the
  identification of the shortest and longest wavelength lines detected
  in each order of the spectrum of \tsa\ is
  indicated.\label{fig:spec_co}}
\end{figure}

A lower resolution $K$ band spectrum has already revealed many
photospheric features from \tsb\ (Duch\^ene et al. 2002). Our new,
higher resolution spectrum also reveals the $^{12}$CO bandheads from
this object, as well as a suite of individual transitions between them
(see Figure\,\ref{fig:spec_co}). Because our radial velocity standard
giant template is of too early spectral type with respect to \tsb, its
spectrum contains less photospheric features, which have different
strengths than that of later type dwarves (Wallace \& Hinkle 1996). In
fact, the CO bandheads are the only significant features that are in
common to both spectra. Conveniently, M0--M1 dwarves and G8--K0 giants
show CO bandheads of essentially the same strength. We can therefore
compare the strength of these features in \tsb\ and HD\,35410
(spectral types M0.5 and G9III, respectively) to estimate the amount
of veiling in the extincted TTS. To match the shape of the bandheads
in the two objects, we needed to convolve the spectrum of the slowly
rotating giant by a 6\,pixel FWHM Gaussian profile, suggesting a
rotational velocity of $v \sin i = 14\pm3$\,km~s$^{-1}$ for
\tsb. After this convolution, we find the bandheads in \tsb\ to be
about three times too weak, which implies that the veiling in Dec 2003
was $r_K\sim2$, similar to its level in Nov 2002. This could be
related to the negligible changes observed for the accretion rate on
this source.

Besides the molecular and atomic emission lines (Sect.~4.1 and 4.2)
and gaseous CO absorption lines (Sect.~4.4), the spectrum of \tsa\ is
remarkably featureless in all orders, despite its high signal-to-noise
ratio. The tentative, very weak $v=2$--0 rovibrational $^{12}$CO
bandhead at 2.29\,$\mu$m in the spectrum of \tsa\ is most likely an
artifact due to contamination by the spectrum of \tsb. Indeed, if this
feature was real, then the other bandheads should be detected at
roughly similar strengths, which is not the case. We therefore
conclude that no photospheric feature typical of late-type stars has
been detected in the spectrum of \tsa\ despite our use of a much
higher resolution than ever before for this source.


\subsection{Gas-phase CO Absorption Lines}

While the spectrum of \tsa\ in orders \#4 and \#5 reveals no
significant $^{12}$CO bandheads, it shows a series of well-defined
narrow absorption lines (Figure\,\ref{fig:spec_co}). These features,
which are not detected in \tsb\ and do not line up with telluric
absorption features, cannot be the result of an improper telluric
correction. Therefore, they are intrinsic to \tsa. Their non-detection
in our previous lower resolution spectrum of this object (Duch\^ene et
al. 2002) can be well explained by their weakness and narrowness.

Comparing the spectrum of \tsa\ with that of HD\,35410, it appears
that all absorption lines detected in the former are also present in
the latter. Consequently, they are $^{12}$CO transitions. Indeed, we
could identify all of them using the high-spectral resolution atlas of
Wallace \& Hinkle (1996). They correspond to the $R$ ($\Delta J=+1$,
order \#4) and $P$ ($\Delta J=-1$, order \#5) $v=2$--0 overtone
rovibrational transitions of $^{12}$CO with $8<J<21$. In normal
photospheres all these lines have essentially the same strength due to
saturation effects and the series extend bluewards towards the
2.293\,$\mu$m bandhead. The observed strong dependence of the line
strength on the $J$ quantum number of the initial state indicates that
we are detecting absorption from a gaseous component that is much
cooler than any photosphere ($T\lesssim1500$\,K).

\begin{figure}
\epsscale{1.0}
\plotone{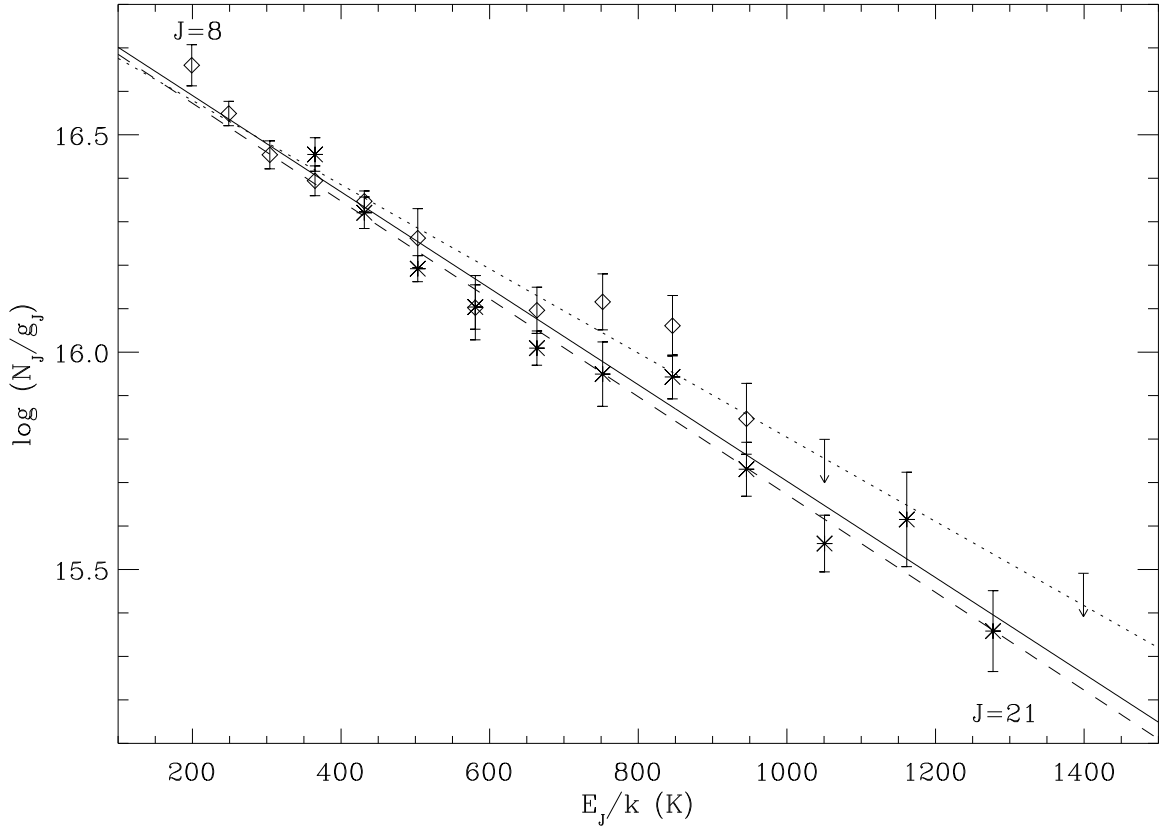}
\caption{Rotational diagram for the $^{12}$CO $P$- (diamonds) and
  $R$-lines (stars) in the spectrum of \tsa.  The dotted and dashed
  lines are linear fit to the $P$- and $R$-lines, respectively while
  the solid line (which corresponds to a gas temperature of 390\,K and
  a total column density $N_{CO}=9.0\times 10^{18}$\,cm$^{-2}$) is a
  simultaneous fit to all detected lines.\label{fig:rot_diag}}
\end{figure}

The observed properties of the absorption lines are presented in
Table\,\ref{tab:co_lines}, together with upper limits for the first
undetected transitions. To derive the properties (temperature, column
density) of the absorbing material, we construct a rotational diagram
for the $^{12}$CO molecule following the method described by Evans et
al. (1991); we use the spontaneous emission coefficients listed in the
HITRAN database (Rothman et al. 2003). We assume that the lines are
optically thin to transform the measured equivalent widths into a
column density of molecules in the state defined by a given $J$
quantum number. The resulting rotation diagram is presented in
Figure\,\ref{fig:rot_diag}. Satisfyingly, the column densities derived
for the $R$ and $P$ lines of the same initial $J$ level agree within
the errorbars.

For optically thin gas at the local thermal equilibrium (LTE), the
molecules should follow a Boltzman distribution. In the
semi-logarithmic rotational diagram, all transitions should therefore
be fit by a straight line, the inverse slope of which is the gas
temperature. As shown in Figure\,\ref{fig:rot_diag}, the measurements
for all $^{12}$CO lines align well along a single straight line for
\tsa. We take this as a proof that our assumptions of the lines being
optically thin and the levels LTE-populated are correct. The first
assumption is satisfied for $N_{CO}\lesssim 5\times 10^{19}$\,
cm$^{-2}$ whereas the LTE assumption implies total gas densities
larger than about 10$^6$\,cm$^{-3}$. Fitting straight lines through
the $R$ and $P$ lines provides fits that are within 1.5$\sigma$ of
each other in both slope and intercept. We therefore proceed to fit a
single straight line to all measured lines simultaneously and derive a
gas temperature of $T_{gas}=390\pm15$\,K.

From the intercept, we derive a $^{12}$CO column density of
9.0$\pm$1.8$\times10^{18}$\,cm$^{-2}$, in agreement with the optically
thin assumption and well above the minimum column density required for
dust- and self-shielding of the CO molecules against photodissociation
by ultraviolet photons
($N_{CO}^{shield}\sim4\times10^{17}$\,cm$^{-2}$, Hollenbach \& Tielens
1997). Since this gas is constrained within a $\lesssim$3\,AU-radius
disk (see Sect.~5.1), this implies a lower limit to the CO mass of
1.3$\pm$0.3$\times10^{-9}\,M_\odot$. Likely, the gas mass is actually
larger given that our absorption measurement only intercepts a
fraction of the disk because of its inclination to our line of
sight. Using a typical 10$^{-4}$ fractional abundance of $^{12}$CO
with respect to H$_2$ (e.g., Frerking, Langer \& Wilson 1982), this
implies a total gas mass that is larger than
9.4$\pm$1.9$\times10^{-7}\,M_\odot$; note, however, that the actual
abundance of CO in this system is unknown. Furthermore, the upper
limit on the disk radius results in a lower limit to the gas density
on order of 2.0$\pm$0.4$\times10^9$\,cm$^{-3}$, in agreement with our
LTE hypothesis. Assuming an interstellar-like gas-to-dust ratio, the
amount of CO gas derived here corresponds to an extinction of
$A_V\sim90$\,mag with the caveat that the gas-to-dust ratio could be
much different value from its interstellar value. As discussed in
Sect.~5.1, we believe that this warm gaseous component traces a
different circumstellar component than the dusty material responsible
for the reddening of the \ts\ system.

The lower excitation $^{12}$CO absorption lines ($J<8$) fall in
between orders \#4 and \#5, so we cannot study them. A much colder
gaseous component could therefore be present in our line-of-sight to
\tsa, as observed towards many high-mass young stellar objects (e.g.,
Mitchell et al. 1990), without us being able to detect
it. Furthermore, we cannot determine whether ice-phase $^{12}$CO is
also present in the environment of this source, as it would be traced
by a broad feature centered around the $P$1 line (e.g., Boogert,
Hogerheijde \& Blake 2002), which is outside our spectral range.

\begin{figure}
\epsscale{1.0}
\plotone{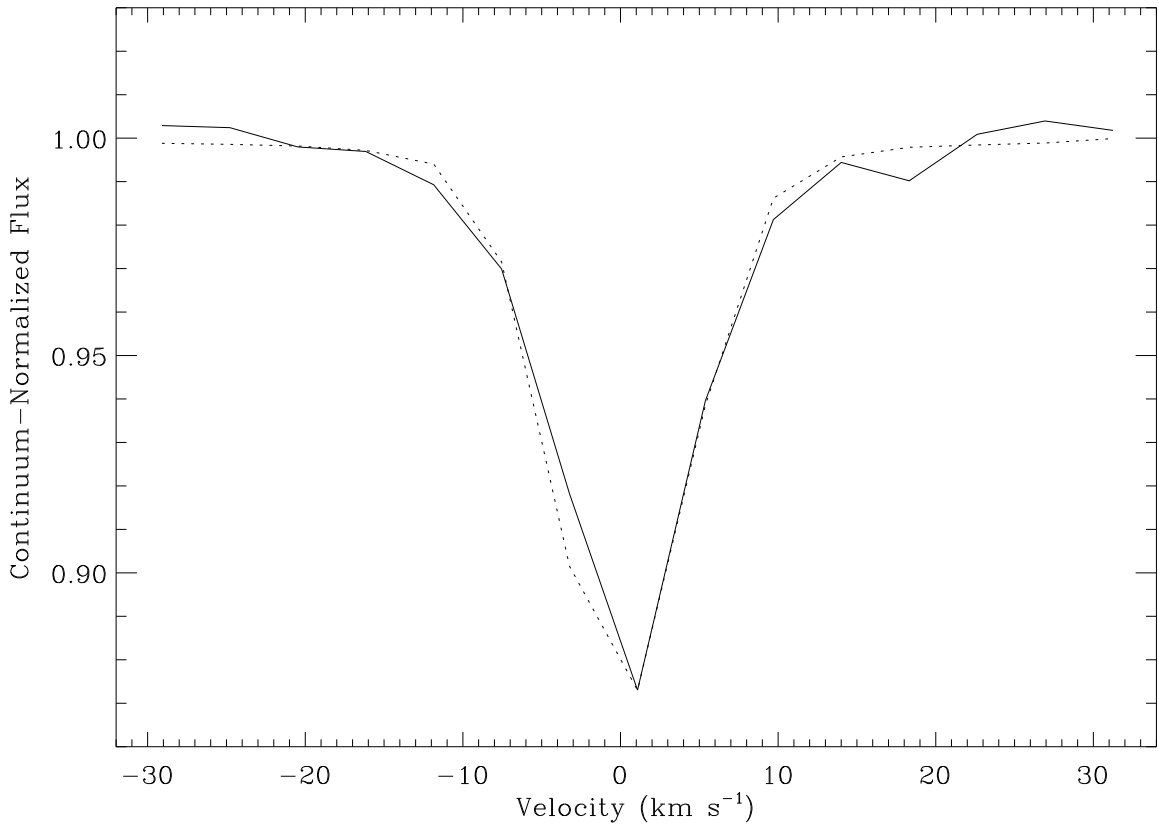}
\caption{The EW-weighted average of the CO absorption line profiles
  (solid line) compared to that of the unresolved arc lamp lines
  (dotted line). Both profiles are averaged over all the features
  found in orders \#4 and 5. The CO lines appear unresolved at our
  spectral resolution of $\sim8.5$\,km~s$^{-1}$.\label{fig:co_prof}}
\end{figure}

Searching for some dynamical information regarding the gas, we compare
the CO absorption features profile with that of the unresolved arc
lamp lines in Figure\,\ref{fig:co_prof}. All detected CO features are
included in the average, in which they are weighted by their
respective EW so that the deeper lines dominate the averaged
profile. As opposed to the 4.7\,$\mu$m CO absorption lines detected
towards several young stellar objects, which frequently show blue-
and/or red-shifted components (Mitchell et al. 1990; Boogert et
al. 2002), the $^{12}$CO lines in our spectrum of \tsa\ are entirely
unresolved and therefore do not show evidence for infall or
outflow. The intrinsic linewidth of the features has to be $\Delta
v<4$\,km~s$^{-1}$ given that the lines do not appear broader than the
unresolved arc lamps lines. Furthermore, no red or blue wing is
detected; the depth of any wing extending beyond 10\,km~s$^{-1}$
cannot be more than $\sim$5\,\% that of the central absorption
feature.

While gaseous $^{12}$CO absorption features have been detected in the
$M$-band spectrum of three embedded young stellar objects (Mitchell et
al. 1990; Boogert et al. 2002; Brittain et al. 2005), they are usually
seen in emission in the spectrum of optically detected TTS and Herbig
AeBe stars (Carr, Mathieu \& Najita 2001; Brittain et al. 2003; Rettig
et al. 2004; Thi et al. 2005). The warm, close circumstellar material
traced by these emission features is seen directly by the observer
thanks to the low-to-moderate inclination of these objects. More
embedded sources and objects observed at high inclination provide a
configuration in which CO lines are found in absorption that is much
more sensitive to detect the gas. To the best of our knowledge, \tsa\
is the second young stellar object (with HL\,Tau, Brittain et
al. 2005) in which gas absorption features are detected at
2.3\,$\mu$m, despite the much weaker strength of the overtone
transitions with respect to the fundamental transitions at
4.7\,$\mu$m, and the first in which the absorbing gas is detected
within a beam size of only 0\farcs05, or a mere 7\,AU at the distance
of \ts.

Encouraged by our detection of gaseous CO absorption features, we
investigated the possible presence of NH$_3$ and CH$_4$ absorption
features. We failed to detect the rovibrational absorption lines from
the ground-state of NH$_{3}$ at 2.24719\,$\mu$m (order \#3) and
CH$_{4}$ at 2.3075\,$\mu$m (order \#4) down to an equivalent width of
0.02\,\AA\ (2$\sigma$).  Assuming the same gas temperature than that
derived for CO, this implies upper limits on the column densities of
about $5\times 10^{17}$ cm$^{-2}$, equivalent to upper limits on the
abundance of these two molecules of about $5\times 10^{-6}$.


\subsection{Radial Velocities}

\begin{figure}
\epsscale{1.0}
\plotone{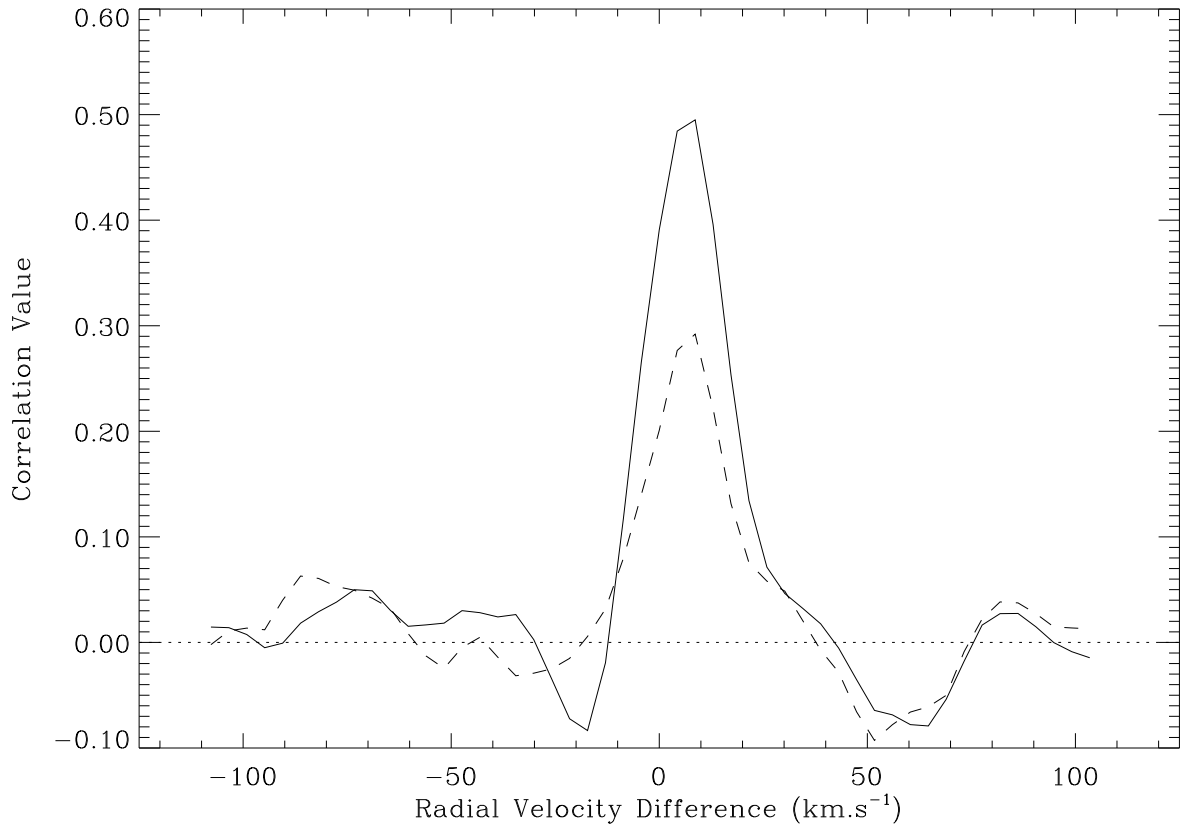}
\caption{Cross-correlation functions for \tsa\ (dashed curve) and \tsb\
  (solid curve) using HD\,35410 as a template. The spectra from orders
  \#4 and \#5 were used simultaneously to obtain the cross-correlation
  functions. Both peaks are highly significant and no other peak of
  similar amplitude is detected even at high velocity shifts for
  either object.\label{fig:correl}}
\end{figure}

The clear detection of photospheric features in the spectrum of \tsb,
combined with the high spectral resolution we achieved, allows to
determine the radial velocity of that component. From the
cross-correlation of \tsb's spectrum with that of HD\,35410 (see
Figure\,\ref{fig:correl}), we find that the observed difference in
radial velocity is 7.3\,km~s$^{-1}$. Taking into account the motion of
the observatory with respect to the Sun, we find that \tsb\ has a
heliocentric radial velocity of $+21.1\pm1.0\pm0.7$\,km~s$^{-1}$. The
latter uncertainty is the systematic error introduced by the
uncertainty on the velocity of the radial velocity standard and the
former is the uncertainty in determining the centroid of the
cross-correlation peak.

We can also use the $^{12}$CO absorption features observed in the
spectrum of \tsa\ to determine a radial velocity. Again, a strongly
significant peak is found in the cross-correlation between \tsa\ and
HD\,35410(see Figure\,\ref{fig:correl}). We find a radial velocity of
$+22.0 \pm 1.0 \pm 0.7$\,km~s$^{-1}$, a velocity that is not
significantly different from that of \tsb. This does not necessarily
represent the radial velocity of the central object; rather it traces
the velocity of material that is located in front of the star. Yet,
this material is likely located in a small circumstellar disk
surrounding \tsa\ (Sect.~5.1), and this radial velocity would then
apply to the central object as well.

For comparison, the radial velocity of the optically bright \tn\ has
been measured to be $+19.1\pm1.2$\,km~s$^{-1}$ (Hartmann et al. 1986).
Therefore, both \tsa\ and \tsb\ are shifted by 2--3\,km~s$^{-1}$ with
respect to \tn, a shift roughly twice as small as the observed motion
of the \ts\ system with respect to \tn\ in the plane of the sky
($\sim5$\,km~s$^{-1}$, Ghez et al. 1995; Roddier et al. 2000; Beck et
al. 2004). Finally, we note that the radial velocity difference
between \tsa\ and \tsb\ is only about 1\,km~s$^{-1}$, whereas the
typical orbital velocity of the system projected in the plane of the
sky is much larger, above 10\,km~s$^{-1}$ (Duch\^ene et al. 2002; Beck
et al. 2004). This implies that the orbit of \tsb\ around \tsa\ is
essentially in the plane of the sky and/or \tsb\ is currently close to
one of its two turnaround points as seen from the Sun.


\section{Discussion}

\subsection{\tsa: An Intermediate Mass Star With an Edge-On Disk?}


\subsubsection{Location Of the Gas Around \tsa}

As demonstrated in Sect.~3.1, \tsb\ suffers a roughly constant
extinction $A_V\sim15$\,mag. Presumably, the material that is
obscuring this component also lies in our line of sight to \tsa, since
this component is also highly reddened. We discuss this further in
Sect~5.2. For now, we focus on \tsa, which we have shown behaves in a
much more dramatic fashion and presents peculiar properties that have
lead to its IRC classification.

The most important result of our high spectral resolution study is the
discovery of the presence of warm CO gas in our line of sight to
\tsa. Because these absorption features are not detected in front of
\tsb\ and because the derived gas temperature is as high as
$\sim$390\,K, this gas has to be located within only a few AU of the
central source. It is therefore part of the circum{\it stellar}
material of \tsa. The fact that the CO absorption features are very
narrow ($\Delta v<4$\,km~s$^{-1}$) implies a dynamically stable
configuration, with no significant infalling or outflowing motion. A
compact envelope around \tsa\ would evolve on a short dynamical
timescale and should show large positive or negative radial velocity
(depending on whether it is in infall or outflow motion) since the
free-fall velocity at 5\,AU of a low-mass star is larger than
10\,km~s$^{-1}$. Therefore, the circumstellar material of \tsa\ cannot
be distributed in a roughly spherical geometry.

The simplest geometry to maintain the gas at a single radial velocity
is an edge-on circumstellar disk, in which the gas intercepted by our
line of sight is exclusively in tangential Keplerian motion; namely
such material presents no radial velocity motion\footnote{Given the
inferred small size of the disk, an ``infalling disk'' similar to that
modeled by Hogerheijde (2001) for L1489\,IRS is excluded here as the
infal motion would be detectable in our line profile (see Boogert et
al. 2002).}. If the collimated jet that lies in the plane of the sky
indeed arises from \ts\ (Solf \& B\"ohm 1999), the fact that \tsa\ may
be surrounded by an edge-on disk appears as no surprise. The small
size of this disk is naturally explained by the disruptive tidal
forces exerted by \tsb\ through its orbital motion (e.g., Artymowicz
\& Lubow 1994). The actual radius of the disk is likely to be on order
2--3\,AU given the binary periastron distance (Johnston et al. 2004;
Beck et al. 2004). The orbital period at the outer edge of the disk is
on order of a few years at most, and our line of sight intercepts a
different section of the disk within just a few weeks. The variability
could therefore be explained by asymmetries within the disk, with
alternatively thicker and thinner parts that move in front of us and
away as the disk rotates. Given its size, this clearly is a different
structure than the much larger absorbing screen found by Walter et
al. (2003) and that we discuss in Sect.~5.2. Note that, since both
\tn\ and \tsb\ posses their own circumstellar disks but are not in an
edge-on configuration, the three disks in this system are not parallel
to each other, as already found other T\,Tauri multiple systems (e.g.,
Stapelfeldt et al. 1998b).


\subsubsection{Nature of \tsa}

If there is indeed a small edge-on disk around \tsa, can it explain
the peculiar properties of this IRC? In particular, the featureless
near-infrared spectrum, the strong variability in near-infrared
brightness and color on a timescale of a few weeks only, the variable
Br$\gamma$ EW tracing the accretion phenomenon, and the high linear
polarization rate derived by Kobayashi et al. (1997) must be accounted
for. If the optical depth of the disk through its midplane is large
($\tau\gtrsim10$), then we can only receive scattered light off the
outer surface of the disk, as in other T Tauri edge-on disk systems
(e.g., Burrows et al. 1996). On the other hand, if the disk opacity is
small, then we receive transmitted light from both the central star
and the inner rim of the disk, attenuated by the disk
self-absorption. The high linear polarization rate favors a pure
scattering regime, but none of the known edge-on disk shows as large a
photometric variability as \tsa\ does. A finer analysis is therefore
required to derive a consistent model for this source.

When \tsa\ fades, it also becomes much redder. This behavior, also
pointed out by Beck et al. (2004) for the unresolved \ts\ system over
a period during which the IRC probably dominated the system in the
near-infrared, suggests a change in obscuration as the main cause of
the source variability. The large color change experienced by \tsa\
when its flux varies shows that scattering is not the dominant
phenomenon, as the latter introduces no time-dependent color
changes. We therefore suggest that the edge-on disk surrounding \tsa\
is only slightly or moderately opaque, with a near-infrared optical
depth $\tau_K$ no larger than a few when the object is brightest. This
moderate opacity ensures that scattering does not dominate in the
bright state while changes of opacity of a few remain reasonable: the
observed 3\,magnitude drop in $K$ band brightness for \tsa\ implies an
increase in opacity by $\Delta\tau_K\sim 3$ at that wavelength. Yet,
with an opacity of a few in the near-infrared, scattering may account
for a non-negligible fraction of the object's total flux, thereby
leading to a linear polarization rate of a few percent, as observed by
Kobayashi et al. (1997). We note that the column density of warm CO we
have derived implies $\tau_K\sim9$ assuming interstellar gas-to-dust
and $^{12}$CO/H$_2$ abundance ratios. While this may be slightly too
large, this is nonetheless in good agreement with the idea that the
disk is only moderately optically thick.

The $N$ band variability does not follow this ``redder when fainter''
trend, as already suggested by the multi-wavelength amplitude of the
1990 flare: Ghez et al. (1991) found that $\sim2$\,mag $N$ band
variations accompanied an equal amplitude $K$ band flare whereas
$\tau_N/\tau_K\lesssim0.5$ for interstellar matter (Rieke \& Lebofsky
1985). Furthermore, comparing a recent November 1999 $N$ band
measurement (McCabe 2004; McCabe et al. 2005, in prep.) to the late
1990 measurements of Ghez et al. (1991), we find that between the two
observations, \ts\ has brightened by $\sim0.3$\,mag in $K$ (Beck et
al. 2004) whereas it has {\it faded} by $\sim0.5$\,mag in $N$. In both
epochs, \ts\ was near its maximum brightness and, therefore,
presumably dominated by \tsa. Clearly, variable obscuration cannot
account alone for the variability of \tsa. We propose that the
near-infrared light comes from the central source and/or the innermost
regions of the circumstellar disk, and is therefore partially absorbed
by the outer parts of the disk whereas the 10\,$\mu$m emission comes
from further out, close to the outer edge of the small disk. This way,
the fluctuations of both wavelengths may not be
correlated. Variability in the $N$ band would arise from a changing
amount of material located toward the observer beyond $\sim$1\,AU of
the central star, possibly because of outer spiral structures
triggered by the orbital motion of \tsb, for instance.

With the proposed geometry for \tsa, the light we receive is a
combination of starlight and inner disk emission, both of them
suffering from absorption by the disk itself. The absence of low-mass
star photospheric features in the spectrum of \tsa\ either means
that the central star is intrinsically featureless, implying a
spectral type between late B and mid-F or so, or that the disk
emission is much larger than that of the star itself. In the latter
case, the object would be a FU\,Ori-like object, but our spectrum does
not show the rotationally broadened $^{12}$CO bandheads typical of
these objects (Hartmann \& Kenyon 1996), making this interpretation
unlikely. We therefore favor an intermediate-to-early spectral type
for \tsa, making it the earliest type and consequently highest mass
component of the T\,Tau triple system.

In its historically brightest state, in late 1999, \tsa\ was 0.5\,mag
fainter at $K$ than \tn\ (Beck et al. 2004), which in fact implies
that \tsa\ was the brightest component of the system by about 1\,mag
once the $A_V\sim15$\,mag extinction screen in front of \ts\ is taken
into account\footnote{Rigorously, the $A_V$ estimate and brightest $K$
magnitude are not simultaneous. However, at the time of our first
$A_V$ estimate, in late 2000, \tsa\ was only 0.8\,mag fainter than
\tn\ at $K$. At that time, the dereddened flux of \tsa\ was therefore
still larger than that of \tn\ by a factor of $\sim2$.}. Despite its
non-detection at visible wavelengths, \tsa\ is therefore intrinsically
very bright in the near-infrared, at least at times, especially when
one considers that the extinction to \tsa\ is likely to be larger than
that to \tsb. In fact, at the epochs when \tsa\ was as bright as
$K\lesssim5$ (dereddened), it was the second brightest near-infrared
source in the entire Taurus star-forming region after AB\,Aur, a
80\,$L_\odot$ B9 Herbig Be star (Kenyon \& Hartmann 1995). In other
words, \tsa\ is most likely one of the highest mass object in the
Taurus star-forming region even though it is not optically detected.

From this set of converging pieces of evidence, we suggest that \tsa\
is an intermediate mass star with a 2--3\,AU-radius edge-on disk that
is partially transparent in the near-infrared. A 2.5--3\,$M_\odot$
stellar mass would be consistent with an early spectral type as well
as a bright intrinsic near-infrared brightness based on the observed
properties of AB\,Aur. The current lower limit to the dynamical mass
of the system (Beck et al. 2004) is also consistent with such a mass
estimate for \tsa. Another consistent argument is the gas temperature
we derived in Sect.~4.4: the temperature resulting from direct
illumination by a 9000K photosphere is on order 270\,K at 3\,AU. An
intermediate mass star would have a luminosity of several tens of
$L_\odot$, much larger than the bolometric luminosity of \tsa. This
apparent light deficit could be explained by a combination of two
factors. First, the edge-on disk scatters a large fraction of the
optical and near-infrared starlight away from our line of sight. In
the case of HH\,30, Wood et al. (2002) have shown that only 10\,\% of
the intrinsic starlight reaches the observer, resulting in a largely
underestimated bolometric luminosity. The disk surrounding \tsa\ is
however only partially opaque, and so the luminosity loss is probably
not as dramatic. The second factor is the presence of a foreground
absorbing screen which also scatters light away from the observer
(cf. Sect.~5.2). Either way, the basic argument is that the
extinguishing material around \tsa\ is not spherically symmetric and
therefore our bolometric luminosity estimate, largely driven by the
unextincted mid-infrared photometry, is only a lower limit to the
object's actual luminosity.


\subsubsection{A Plausible Physical And Chemical Model Of The Disk
  Around \tsa}

The derived temperature ($\sim 390$\,K) and density ($\sim 2\times
10^{9}$\,cm$^{-3}$) allow to estimate the disk mass, when they are
compared with predictions of the disk structure. Indeed, the derived
temperature and density compare remarkably well with the disk
structure predicted by the model by Dullemond \& Dominik
(2004). Assuming that the disk is illuminated by a 80\,$L_{\odot}$
central star, following the arguments presented above, a disk with a
radius of 3 AU and a mass of 0.003 M$_{\odot}$ would have a warm layer
of CO at about the observed density and temperature (C. Dominik,
private communication). The disk would be seen at an inclination angle
of about 80\degr, i.e., almost but not exactly edge-on. The disk mass
inferred from this model is much larger than the value we have derived
from the CO column density because we do not intercept the disk
midplane, which would be markedly cooler (around 150\,K), and, hence,
we do not intercept the bulk of the disk mass with our
observations. However, the inferred disk mass is satisfyingly lower
than the upper limit on the \ts\ disk mass derived by Akeson et
al. (1998). In practice, this comparison with a theoretical structure
suggests that the disk is not seen exactly edge-on but through
material located at moderate elevation above the disk midplane. This
makes \tsa\ a rare and precious source where absorption studies of the
molecular content can be carried out, because the disk is not exactly
edge-on, which would result in all received photons being scattered
rather than transmitted through the disk. A more detailed physical
model of the disk surrounding \tsa\ is not within the scope of the
present study but would greatly help refining our understanding of
this object.

The upper limits on the ammonia and methane abundances ($\leq 5\times
10^{-6}$) are relatively stringent when compared to theoretical
expectations. Both ammonia and methane are believed to be formed on
the grain surfaces, by active grain chemistry (Tielens \& Hagen 1982),
and have indeed been observed in the solid form with abundances around
$10^{-6}-10^{-5}$ (e.g. Boogert et al. 1996; Dartois et al. 1998;
Alexander et al. 2003). In the region where we observed warm CO, the
dust temperature is certainly large enough to make the grain mantles
sublimate. Therefore, the mantle components are injected into the gas
phase, and ammonia and methane are no exception. For this reason, one
would expect that the same amount of ammonia and methane seen in the
ices are found in the gas phase. This is the case, for example, of
ammonia whose gas phase has been measured to be of order of
$10^{-6}-10^{-5}$ in the hot cores of massive protostars and in the
protostellar outflows (e.g., Bachiller 1996; Krutz et al. 2000). The
same applies to methane, whose gas phase abundance in the massive hot
cores has been measured to be more than $10^{-6}$ (e.g. Boogert, Blake
\& Oberg 2004). So, why we do not detect ammonia and methane at a
level larger than $5\times 10^{-6}$? There are two possibilities. It
is possible that the grain mantles around \tsa\ have a different
composition than the other mentioned environments, and are less
enriched of ammonia and methane. Alternatively, chemical reactions
could have removed the two molecules to form more complex molecules,
which is expected in some cases in the hot cores (e.g. Charnley,
Tielens \& Millar 1992). Both explanations are plausible, and would
have, if confirmed, interesting implications. For example, the
abundance of ammonia in ices traces back to the conditions where the
ices formed, namely during the pre-collapse phase, unless a vigorous
reprocessing occurred during the disk phase. Further, and more
sensitive observations are in need to fully explore these
possibilities and the linked consequences.


\subsection{On The Environment Of The \ts\ Binary System}

We have found \tsb\ to be extincted by about $A_V\sim15$\,mag at the
time of our observations, a value that changed little over the last
few years. This number is similar to the extinction derived from the
ice absorption feature by Beck et al. (2004), suggesting that both
approaches probe the same physical structure. Estimating the depth of
the mid-infrared silicate absorption feature from three narrow-band
photometric measurements, Ghez et al. (1991) estimated $A_V\sim5$\,mag
only. More recently, Herbst, Robberto \& Beckwith (1997) found the
silicate feature to be almost twice as deep, bringing this extinction
estimate closer to the other ones. We note, however, that the depth of
the 10\,$\mu$m silicate feature only poorly correlates with $A_V$ in
molecular clouds (e.g., Whittet et al. 1988). Still, a consistent
picture of the system can be proposed based on the convergent
estimates of the amount of extincting gaseous and solid materials. As
readily suggested by the non-detection of either component at visible
wavelengths, there seems to be an $A_V\sim15$\,mag obscuring cloud in
front of the entire \ts\ system, which we believe to be the structure
seen in absorption by Walter et al. (2003). This structure, whose size
is about 0\farcs7$\times$0\farcs5, is much too large to correspond to
a circumstellar disk around either star given their small
separation. It is quite possible that this is a circumbinary envelope
or thick disk ({\it \`a la} GG\,Tau, Guilloteau, Dutrey \& Simon 1999)
that obscure both components. If its inner radius is at least 50\,AU
(or $\sim$0\farcs35), it is stable with respect to the orbital motion
of the inner tight binary system whereas its outer radius is probably
set by the motion of \tn, which is located far away enough to generate
only little perturbation to such a structure (Walter et
al. 2003). Such a large size for the obscuring screen is also
consistent with the presence of ice and silicate absorption features:
only that far from the central star is the material cold enough to
produce significant absorption in these features. On the other hand,
the warm CO absorption features we have found in the spectrum of \tsa\
cannot be associated with this screen, since the gas temperature
50\,AU from a source with a few times 10\,$L_\odot$ luminosity is not
higher than $\sim120$\,K.

We have shown that the molecular hydrogen emission from \ts\ is
spatially resolved over 2\arcsec\ around the system and does not arise
from the stars themselves. The linear accelerations observed in the
position-velocity diagram are reminiscent of propagating jets (e.g.,
Hirth, Mundt \& Solf 1997); however, the presence of blue-shifted
components {\it on both sides} of the system and the narrowness of the
line profiles at any spatial location argues against the emission line
being excited in shocked regions of a single jet. On the other hand,
the spatial coincidence between ultraviolet fluorescent line emission
with the infrared line emission we detect to the West of the system
argues for a similar emission mechanism; however, no ultraviolet
emission seems to be directly associated with the infrared emission we
detect East of the tight binary system. It is not possible to conclude
on the nature of the excitation mechanism on the basis of our
observations only; similar observations with different slit
orientations, or integral field spectroscopic observations, would be
required to determine the exact nature of the H$_2$ emission. In any
case, our observations reveal the complex kinematic and spatial
structure of the gas surrounding \ts.


\section{Conclusion}

We have used the AO-fed cross-dispersed echelle spectrograph NIRSPEC on
Keck~II to obtain the first high spatial (0\farcs05) and spectral
($R\sim35000$) resolution 2\,$\mu$m view of the mysterious tight
binary system \ts\ by aligning the entry slit along the position
angle of the binary. We have further obtained the first 3.8 and
4.7\,$\mu$m broadband images that resolve all three components of the
T\,Tau multiple system, as well as new 1.6 and 2.2\,$\mu$m images.

The spectrum of \tsb\ confirms that it is a low-mass TTS with
significant excess emission from its circumstellar disk. Its
very red near-infrared colors can be explained by a roughly constant
extinction on order $A_V\sim15$\,mag, in agreement with previous
extinction measurements based on ice and silicate features. We believe
that the obscuring material is located in a $\gtrsim$50\,AU-sized
circumbinary structure, whose absorption was also recently detected in
the ultraviolet.

The spectrum of \tsa, on the other hand, is featureless, with the
notable exceptions of i) a weak and variable Br$\gamma$ emission that
probably traces accretion on the central star, and ii) a series of
narrow $^{12}$CO rovibrational absorption lines without their
corresponding bandhead. A rotational diagram shows that the CO lines
correspond to a moderate column density of gas at a temperature on
order 390\,K. To account for this high temperature and in the absence
of evidence for infall or outflow in the gas, we believe that this
material is located in a small (2--3\,AU in radius) edge-on disk that
surround an intermediate-mass star. The large variability of \tsa\ in
both near-infrared brightness and color cannot be uniquely accounted
for by changes in the amount of line of sight extinction. Rather, we
propose that the disk around \tsa\ is moderately opaque
($\tau_K\sim$\,a few) and that our line of sight intercept
alternatively thicker and thinner sections of the disks as its outer
radius rotates around the central star in just a few years.

Finally, we have analyzed the spatial and kinematic properties of the
molecular gas in the vicinity of \ts, as traced by molecular hydrogen
emission. We find an unusual structure in that the emission appears to
gradually blueshift with distance to the stars on both sides of the
binary. The exact nature of the excitation mechanism remains unknown
but these results confirm the highly complex structure of the gaseous
material surrounding the T\,Tau system.

\acknowledgments

We are grateful to our referee, T. Beck, for her detailed report that
helped us clarifying several aspects of this manuscript and to
C. Dominik for granting us access to unpublished details of his disk
models. Data presented herein were obtained at the W. M. Keck
Observatory, which is operated as a scientific partnership among the
California Institute of Technology, the University of California, and
the National Aeronautics and Space Administration. The Observatory was
made possible by the generous financial support of the W. M. Keck
Foundation. This work has been supported in part by the National
Science Foundation Science and Technology Center for Adaptive Optics,
managed by the University of California at Santa Cruz under
cooperative agreement AST 98-76783 and by the Packard Foundation. The
authors wish to extend special thanks to those of Hawaiian ancestry on
whose sacred mountains we are privileged to be guests. Without their
generous hospitality, none of the observations presented herein would
have been possible.


{}

\clearpage

\begin{deluxetable}{ccccc}
\tablecaption{New astrometric measurements in the
  T\,Tau triple system\label{tab:astrom}}
\startdata
\tableline
 & \multicolumn{2}{c}{\tn--\tsa} & \multicolumn{2}{c}{\tsa--\tsb} \\
Date & Sep. (\arcsec) & Pos. Ang. (\degr) & Sep. (\arcsec) &
  Pos. Ang. (\degr) \\
\tableline
2002 Dec 13 & 0\farcs695$\pm$0\farcs007 & 183\fdg3$\pm$0\fdg7 &
  0\farcs108$\pm$0\farcs001 & 284\fdg9$\pm$0\fdg9 \\
2003 Dec 12 & 0\farcs698$\pm$0\farcs005 & 181\fdg9$\pm$1\fdg2 &
  0\farcs118$\pm$0\farcs002 & 288\fdg6$\pm$1\fdg1 \\
\enddata
\end{deluxetable}

\begin{deluxetable}{ccccc}
\tablecaption{New photometric measurements in
  the T\,Tau triple system\label{tab:phot}}
\startdata
\tableline
Date & Filter & \tn\tablenotemark{a} & \tsa & \tsb \\
\tableline
2002 Dec 13 & $K_{cont}$ & 5.52 & 8.77$\pm$0.03 &
  8.33$\pm$0.03 \\ 
 & Br$\gamma$ & 5.52 & 9.10$\pm$0.03 & 8.48$\pm$0.03 \\
 & $L'$ & 4.32 & 5.64$\pm$0.03 & 6.25$\pm$0.03 \\
 & $M_s$ & 2.95 & 3.78$\pm$0.03 & 4.59$\pm$0.05 \\
2003 Dec 12 & $K$ & 5.52 & 9.70$\pm$0.07 & 8.50$\pm$0.06 \\
 & $H$ & 6.32 & $>$13.6 & 10.22$\pm$0.11 \\ 
\enddata 
\tablecomments{All measurements are presented as magnitudes.}
\tablenotetext{a}{The photometry for \tn\ is taken from Beck et
al. (2004) at $K$ (its magnitude in the $K_{cont}$ and Br$\gamma$
narrow band filters is assumed equal to that of the broad band
filter) and $L'$ and from Ghez et al. (1991) at $H$ and $M$; we
 applied the latter flux density for our $M_s$ photometry despite the
 slight bandpass mismatch.}
\end{deluxetable}

\begin{deluxetable}{ccccc}
\tablecaption{Instrumental set-up for the spectroscopic
  observations\label{tab:setup}}
\startdata
\tableline
Order \# & $\lambda_i$ & $\lambda_f$ & $\delta v$ & Note \\
 & ($\mu$m) &  ($\mu$m) & (km~s$^{-1}$ / pixel) & \\
\tableline
1 & 2.102321 & 2.132959 & 4.22 & H$_2$ \\
2 & 2.161989 & 2.193496 & 4.28 & HI Br$\gamma$ \\
3 & 2.225272 & 2.257702 & 4.32 & H$_2$, NH$_3$ \\
4 & 2.292261 & 2.325667 & 4.35 & $^{12}$CO, CH$_4$  \\
5 & 2.363703 & 2.398150 & 4.31 & $^{12}$CO \\ 
\enddata
\tablecomments{$\lambda_i$, $\lambda_f$ and $\delta v$ represent the
wavelength at the first and last pixel and pixel width along the
spectral direction, respectively. The average value of $\delta v$
throughout the five orders is used for wavelength calibration in our
study.}
\end{deluxetable}

\begin{deluxetable}{ccc}
\tablecaption{Properties of the \brg\ emission
  lines\label{tab:h_lines}}
\startdata
\tableline
 & \tsa & \tsb \\
\tableline
EW (\AA) & 0.78$\pm$0.05 & 6.2$\pm$0.3 \\
EW$_{red}$ (\AA) & 0.22$\pm$0.05 & 1.6$\pm$0.1 \\
EW$_{blue}$ (\AA) & 0.56$\pm$0.05 & 4.7$\pm$0.3 \\
FWHM (km~s$^{-1}$) & 164$\pm$17 & 147$\pm$9 \\
FWZI (km~s$^{-1}$) & 275$\pm$45 & 490$\pm$43 \\
$v_{peak}$ (km~s$^{-1}$) & $-$17$\pm$22 & $-$30$\pm$4 \\
$v_{red}$ (km~s$^{-1}$) & $+$91$\pm$13 & $+$177$\pm$9 \\
$v_{blue}$ (km~s$^{-1}$) & $-$185$\pm$13 & $-$315$\pm$40 \\
\enddata
\tablecomments{$EW$ is the total EW of the emission line while
  $EW_{red}$ and $EW_{blue}$ are the EWs of the line on either side of
  the rest velocity of \tsb; FWHM and FWZI are the full width at half
  maximum and at 10\% (``zero'') intensity of the emission line,
  respectively; $v_{peak}$, $v_{red}$ and $v_{blue}$ are the measured
  velocities at the line peak intensity and at the extremity of the
  red and blue wings, respectively. All uncertainties include
  uncertainties in the continuum estimates and in the exact line
  profile in the extended wings. Uncertainties related to the absolute
  wavelength calibration (velocity estimates and ``red'' and ``blue''
  EWs) are not included.}
\end{deluxetable}

\begin{deluxetable}{ccccc}
\tablecaption{$^{12}$CO 2--0 rovibrational absorption line
  properties\label{tab:co_lines}}
\startdata
\tableline
& \multicolumn{2}{c}{$R$ lines ($\Delta J = +1$)} &
  \multicolumn{2}{c}{$P$ lines ($\Delta J = -1$)} \\
$J$ & $\lambda_{obs}$ ($\mu$m) & $EW$ (\AA) &
  $\lambda_{obs}$ ($\mu$m) & $EW$ (\AA) \\
\tableline
 8 & & & 2.36567 & 0.187$\pm$0.020 \\
 9 & & & 2.36825 & 0.158$\pm$0.010 \\
10 & & & 2.37083 & 0.138$\pm$0.010 \\
11 & 2.32519 & 0.150$\pm$0.013 & 2.37345 & 0.130$\pm$0.010 \\
12 & 2.32366 & 0.120$\pm$0.010 & 2.37609 & 0.125$\pm$0.007 \\
13 & 2.32218 & 0.097$\pm$0.007 & 2.37875 & 0.109$\pm$0.017 \\
14 & 2.32071 & 0.085$\pm$0.010 & 2.38144 & 0.080$\pm$0.014 \\
15 & 2.31931 & 0.073$\pm$0.007 & 2.38416 & 0.084$\pm$0.010 \\
16 & 2.31789 & 0.068$\pm$0.012 & 2.38689 & 0.093$\pm$0.014 \\
17 & 2.31657 & 0.072$\pm$0.008 & 2.38967 & 0.086$\pm$0.014 \\
18 & 2.31524 & 0.047$\pm$0.007 & 2.39263 & 0.054$\pm$0.010 \\
19 & 2.31397 & 0.033$\pm$0.005 & & $<0.050$ \\
20 & 2.31276 & 0.040$\pm$0.010 & & \\
21 & 2.31152 & 0.023$\pm$0.005 & & \\
22 & & $<0.033$ & & \\
\enddata
\tablecomments{$J$ represents the rotational quantum number of the
  initial state of the transition. The $R$ lines with $J<11$ and $P$
  lines with $J<8$ are outside of the wavelength range covered in our
  dataset. R and P lines are detected in orders \# 4 and 5,
  respectively. For the first undetected line in each series,
  3$\sigma$ upper limits are given. Similar (or less constraining)
  upper limits apply to the following undetected lines within the
  observed wavelength range.}
\end{deluxetable}


\end{document}